\documentclass{elsart}
\usepackage{graphicx}
\usepackage{amsmath}
\usepackage{amssymb}
\usepackage{color}
\journal{Materials and Something}
\graphicspath{{./figures/}}

\begin{document}
\begin{frontmatter}
\title{Polyconvex Model for Materials with Cubic Anisotropy}
\author{Nayden Kambouchev}, 
\author{Javier Fernandez},
\author{Raul Radovitzky\corauthref{cor}} 
\corauth[cor]{Corresponding
author. Tel.: +1-617-xxx-xxxx; fax: +1-617-xxx-xxxx}
\ead{rapa@mit.edu} 
\address{Department of Aeronautics and
Astronautics, Massachusetts Institute of Technology, Cambridge, MA
02139, USA}
\begin{abstract}

Polyconvexity is one of the known conditions which guarantee existence
of solutions of boundary value problems in finite elasticity.  In this
work we propose a framework for development of polyconvex strain
energy functions for hyperelastic materials with cubic anisotropy.
The anisotropy is captured by a single fourth order structural tensor
for which the minimal polynomial basis is identified and used for the
formulation of the strain energy functions.  After proving the
polyconvexity of some polynomial terms, we proceed to propose a model
based on a simple strain energy function.  We investigate the behavior
of the model analytically in one dimension and numerically in two and
three dimensions.  These investigations allow us conclude that the
model possesses the physically relevant directional properties, in
particular, strains in different directions are different and the lack
of any loading symmetries leads to development of ``shear'' stresses
and displacements.
\end{abstract}
\end{frontmatter}

\section{Introduction}
\label{sec:introduction}

Anisotropic engineering materials exhibit directionality in their
mechanical characteristics even when subjected to very large
strains. These materials represent a wide range of applications in
composites and crystals as well as in bio-mechanical systems. Although
some advances have been made in the direction of characterizing simple
cases of anisotropic material behavior through the proposal of
phenomenological models respecting the applicable mathematical
theories, the field is far from completion. In general most of the
phenomenological studies lack detailed analyses on the mathematical
properties of the proposed models.  Even in \cite{Scho-Neff:03}, where
an analysis of general convexity conditions for transversely isotropic
materials is extensively presented, the relationship between the
numerous proposed functions with their physical counterparts is still
to be fully developed.

The material symmetries of an oriented continuum impose definite
restrictions on the form of constitutive relations. The procedure used
for the construction of constitutive models must from the very
beginning assure that the equations are written in a proper manner
which reflects the material symmetries. Furthermore, the final goal of
the procedure is the development of a mathematical framework that
satisfies conditions guaranteeing the existence of solutions for
models that lack the standard regularity properties assuring existence
and uniqueness. Indeed, uniqueness should not be required because it
precludes description of some important physical effects of great
interest, for example, buckling (in this respect see \cite{Ball:77}).
The procedure, presented in this work, follows the approach laid out
in \cite{Scho-Neff:03}.

The fundamental aim from a mathematical perspective is to guarantee
the existence of solutions. Existence of minimizers of some
variational principles in finite elasticity is based on the concept of
quasiconvexity (introduced by Morrey in \cite{Morrey:52}), which
ensures that the functional to be minimized is weakly lower
semi-continuous. Unfortunately, quasiconvexity gives rise to an
integral inequality, which is extremely difficult to handle due to its
global character.  Therefore we turn to the more practical concept of
polyconvexity (\cite{Ball:77}) which can be verified locally.

The increased complexity of observed mechanical behavior of
anisotropic materials requires invariant formulations of anisotropic
constitutive laws. The theory of tensor function representations
constitutes a rational procedure for consistent mathematical modeling
of complex anisotropic material behavior. A particularly strong push
in that direction is the work \cite{Weiss:96} by Weiss who introduced
an exponential function in terms of the mixed invariants.

As extensively presented in \cite{Scho-Neff:03}, the complex
mechanical behavior of elastic materials with an oriented internal
structure at large strains can be described with tensor-valued
functions of several tensor variables: the deformation gradient and a
few additional structural tensors. The follow-up strategy is to
construct constitutive models with an invariant form of the strain
energy function.  The general forms of tensor-valued functions have
been derived in the form of of representation theorems for tensor
functions (\cite{Weyl:46}). The type and minimal number of the scalar
variables entering the constitutive equations are also known. The
interested reader should consult
\cite{Spencer:71,Boehler:79,Boehler:87,Betten:87,Smith&Rivlin:57,Smith&Rivlin:58}
for details.

In this paper we are concerned with the problem of determining the
general form of scalar-valued polynomial expressions, and for that
reason we make use of the concept of integrity basis. So far in the
literature only the simplest form of material anisotropy represented by
transversely isotropic materials with a single preferred principal
direction, has been extensively developed following these lines. In
this work we develop a procedure for the construction of polyconvex
free energy functions for cubic crystal systems. These cubic crystal
systems present three orthogonal principal directions, giving rise to
a considerable increase in the complication of the mathematical
machinery to be dealt with. The main difference with previous works
comes from the need to use a fourth order structural tensor to
characterize the material symmetry group. The need for the fourth
order tensor comes from our desire to use a single structural
tensor. We will make use of results obtained by Zheng
(\cite{Zheng:93}) on the single structural tensors characterizing the
crystal classes.

To summarize, this work presents a large deformation mathematical
model for anisotropic materials with cubic symmetry.

The paper is organized as follows. In Section \ref{sec:cont_mechanics}
we present the basic notation and and review some kinematics relations
at finite strains to be used in the sequel. After that we focus on the
presentation of the mathematical framework for hyperelastic materials
which guarantee a priori some meaningful physical conditions, in
particular, the material frame indifference and the material symmetry
conditions; it will be shown that these two conditions require the
introduction of the concept of structural tensors. Section
\ref{sec:free_energy} is concerned with the application of the
concepts of hyperelasticity and structural tensors to the particular
case of a material formed by cubic crystal. After characterizing the
material symmetries associated with the cubic crystal anisotropy by
means of the appropriate structural tensor, we present a procedure to
build up free energy functions for cubic crystal materials that
fulfill the appropriate mathematical requirements, more specifically
the polyconvexity condition. The proposed functions have the
invariants of the deformation gradient and the structural tensor as
arguments. This approach requires the concept of polynomial integrity
basis, which is also presented.  The representation for the stresses
and the tangent matrix is given in detail. A model fulfilling all the
requirements mentioned above is finally proposed in section
\ref{sec:model}. Two conditions are added to fully determine the
problem and relate it to the physical data.  These conditions are the
stress-free reference configuration and the linearized behavior near
the natural state ${\bf {C}}={\bf {1}}$; these are dealt with in
section \ref{sec:conditions}. We consider the behavior of the proposed
model in one dimension in the next section, where its physically
desirable stress-strain response can be fully appreciated. A short
summary of the variational and finite element formulation is given in
section \ref{sec:variational}. Section \ref{sec:numerical} presents
numerical results obtained from simulation examples using the proposed
model. Finally, in section \ref{sec:conclusion} the main conclusions
of the present work are summarized. A few appendices have been added
at the end to encompass some of the derivations.

\section{Foundations of continuum mechanics}
\label{sec:cont_mechanics}

In the following we consider the class of hyperelastic materials for
which we postulate the existence of a free energy function.  The
resulting constitutive equations must fulfill some requirements that
naturally arise from physical considerations of response invariance of
the material under arbitrary coordinate system transformations. It
will be shown that the requirement on the constitutive functions of
anisotropic solids to satisfy the material frame indifference will
force these functions to be isotropic tensor functions. Therefore the
material symmetry condition cannot be accomplished simultaneously with
the material frame indifference.  In order to resolve this issue we
will make use of the concept of structural tensors.  Structural
tensors increase the number of arguments of the energy functions,
enabling the model to account separately for the material symmetries.

\subsection{Notation and kinematics}

In this section we briefly present the notation and main results
corresponding to kinematics in the standard theory of continuum
mechanics.  The theory presented here is based on a material
formulation.

The movement of a continuum body can be seen as a family of
configurations ordered by the time parameter. Thus, for every
$t\in[0,T]\subset\Re^{+}$, the application $\phi_{t}:B\rightarrow
S\subset\Re^{3}$ is a deformation which transforms the reference
configuration $B$ in the configuration $S$ at time $t$. Then, ${\bf
{x}}=\phi_{t}({\bf {X}})=\phi({\mathbf{X},t})$ identifies the position
$\mathbf{x}$ of point ${\bf {X}}$ at time $t$. We will follow a
lagrangian description of the motion, which implies that the material
coordinates of a point, $\{ X_{A}\}$, are taken as independent
variables.  It is usually called \emph{material description} of the
motion.  The deformation gradient ${\bf {F}}$ is defined as
\[
{\bf {F}}\equiv\nabla\phi_{t}({\bf {X}})
\] 
with the jacobian $J \equiv \mathrm{det}(\mathbf{F})>0$, defined
positive as a condition to prevent material interpenetration.

Let $\dot{{\bf {F}}}$ denotes the material time derivative of the
deformation gradient. It is identical to the material velocity
gradient, i.e.
\[
\dot{{\bf {F}}}=\frac{\partial{\bf {F}}}{\partial t}=DV_{t} .
\]
The deformation gradient ${\bf {F}}$ can be used to form the right
Cauchy-Green tensor, which corresponds to the chosen strain measure,
i.e.
\begin{equation}
{\bf {C}}={\bf {F}^{T}}{\bf {F}}.
\label{eq:c-def}
\end{equation}
In general spaces all inner products appearing in the former and
latter derivations should be properly constructed taking into account
the corresponding space metric, defined as a symmetric second order
covariant tensor, and denoted by ${\bf {G}}$ for the reference
configuration, and by ${\bf {g}}$ for the deformed configuration. In
this paper we will restrict ourselves to the case of euclidean space
with cartesian coordinates in which the metric tensors become the
identity, and therefore they will not explicitly appear in the
calculations.

\subsection{Hyperelasticity and invariance conditions}

As mentioned in the introduction, we will focus our study on
hyperelastic materials. They are an elastic materials class which
postulates the existence of a stored free energy function $\psi({\bf
{X}},{\bf {F}},\cdot)$.  The energy function $\psi$ depends on the
point in the reference configuration ${\bf {X}}$, the deformation
gradient ${\bf {F}}$, and an additional tensor, which characterizes
the material anisotropy. We will restrict ourselves to perfectly
elastic materials, which means that the internal dissipation $D_{int}$
is zero for every admissible process (see \cite{MARSDEN:83}).
Following a standard argument, the constitutive equations relating the
stresses to the energy function are obtained by evaluation of the
Clausius-Planck inequality
\begin{equation}
D_{int}={\bf {P}}:\dot{{\bf {F}}}-\dot{\psi}=({\bf {P}}-\partial_{{\bf
{F}}}\psi):{\bf {F}}\geq0\Rightarrow{\bf {P}}=\partial_{{\bf
{F}}}\psi
\label{eq:dissipa}
\end{equation} 
where the thermal effects have been neglected and $\mathbf{P}$ is the
first Piola-Kirchhoff stress tensor.

The principle of material frame indifference requires the invariance
of the constitutive equation under rigid body motions superimposed
onto the current configuration. i.e., under the mapping ${\bf
{x}}\mapsto{\bf {Q}}{\bf {x}}$ the condition $\psi({\bf
{F}})=\psi({\bf {Q}}{\bf {F}})$ holds for every ${\bf {Q}}$ in the
special orthogonal group $SO(3)$.

As shown in \cite{Truesdell:65} the requirement that the constitutive
equations fulfill the principle of material objectivity yields the
functional dependence $\psi=\psi({\bf {C}})=\psi({\bf {\bf
{C}(}{F})})$, i.e., every dependency on the gradient ${\bf {F}}$ can
be properly substituted for a dependency on ${\bf {C}}$. Now,
considering the relation between the first and second Piola-Kirchhoff
stress tensors, together with the dependency of the energy function on
$\mathbf{C}$ and expression (\ref{eq:dissipa}), we can obtain the
relation between stresses measured by the second Piola-Kirchhoff
stress tensor and the energy function. Thus, we have
\begin{equation}
{\bf {S}}={\bf {F}}^{-1}{\bf {P}}={\bf {F}}^{-1}\partial_{{\bf
{F}}}\psi={\bf {F}}^{-1} (\partial_{\mathbf{C}} \psi
\partial_{\mathbf{F}} \mathbf{C})
\label{eq:s-intermedi}
\end{equation} 
and considering the symmetry of ${\bf {C}}$, we deduce that
$\partial_{\mathbf{C}} \psi \partial_{{\bf {F}}}{\bf {C}}=2{\bf {F}}
\partial_{\mathbf{C}} \psi$, which carried into the expression for
${\bf {S}}$ above gives
\begin{equation}
{\bf {S}}=2\partial_{{\bf {C}}}\psi .
\label{eq:s-derivpsi}
\end{equation}
The anisotropy of a material can be characterized by the material
symmetry group $G_{M}$ with defined respect to a local reference
configuration. The elements of $G_{M}$ are those transformations ${\bf
{Q}}$ that give an invariant material response, i.e., they are
superimposed rotations and reflections on the reference configuration
which do not influence the behavior of the anisotropic material, thus
\begin{equation}
\left\{ \begin{array}{l} \psi({\bf {\bf {C}}{Q}})=\psi({\bf
{C}})\quad\forall{\bf {Q}}\in G_{M} \\ {\bf {P}}({\bf {\bf
{C}}{Q}})={\bf {P}}({\bf {C}}){\bf {Q}}\quad\forall{\bf {Q}}\in
G_{M} . \end{array}\right.\label{eq:fi-P-invari}
\end{equation}
The conditions (\ref{eq:fi-P-invari}) establish that the function
$\psi$ and the tensor ${\bf {P}}$ are $G_{M}-$in\-va\-riant. In general we
have $G_{M}\subset SO(3)$, so the material symmetry group corresponds
to a subgroup of the whole special orthogonal group $SO(3)$, and only
in the case of an isotropic material both groups coincide. This last
fact gives rise to a problem regarding two conflicting requirements:
from one side the functions should be invariant only under
transformation belonging to $G_{M}$, reflecting the material
anisotropy, and from another side their formulation should be
transformation independent, so that the representation is
coordinate-free, i.e., the ${\bf {Q}}$'s in (\ref{eq:fi-P-invari})
should belong to $SO(3)$. To summarize, the material symmetry
condition requires the use of an isotropic function, but at the same
time that leads to loss of the information concerning the material
anisotropy.

It has to be emphasized, however, that so far we have been
considering only constitutive functions dependent on one argument, the
tensor ${\bf {C}}$, and this points out to one possible approach to
meet both requirements.  It will be shown that both requirements can
be satisfied simultaneously by extending the tensorial argument list
of the energy functions, thus obtaining an isotropic
function embodying the anisotropy information.  This approach is put
into practise by means of \emph{structural tensors}.

\subsection{Isotropic tensor functions for anisotropic material 
response. Structural tensors}

As it has been shown in the previous section, the constitutive
equations can be deduced from a free energy function, but we faced the
problem of characterizing anisotropic materials with a dependency on
${\bf {C}}$ only. We saw that it was not possible to have both the
anisotropy and the invariance under any spatial rotation and reflexion
captured in that manner.  The idea behind the structural tensors is to
be able to have an isotropic tensor function, i.e., one which is
invariant under any rotation in space, but at the
same time to retain the symmetry information characterizing the
anisotropy of the material.  Both conditions of rotation invariance
and anisotropy can be properly fulfilled by adding more tensors as
additional arguments in the free energy function.

The \emph{structural tensors} are useful in obtaining irreducible and
coordinate-free representations for anisotropic tensor functions
because they characterize the spatial symmetry group.  The concept of
structural tensors was introduced by Boehler in \cite{Boehler:79}.
The characterization of the symmetry group is done in the following
sense: the tensors ${\bf {\xi}},...,{\bf {\zeta}}$ are said to be the
structural tensors of the spatial symmetry group $G_M$ if and only if
\[
\left. \begin{array}{l}
Q_{ij}\ldots Q_{kl}\xi_{j\ldots l}=\xi_{i\ldots k}\\
Q_{ij}\ldots Q_{kl}\zeta_{j\ldots l}=\zeta_{i\ldots
  k}\end{array}\right\} \Longleftrightarrow {\bf {Q}}\in G_{M}.
\]
Basically, the effect of the structural tensors can be captured in the
difference between the following two statements
\begin{itemize}
\item The relation $\psi({\bf {C}},{\bf {\xi}})=\psi({\bf {Q}^{T}{\bf {C}}{\bf {Q}},{Q}\star}{\xi})$
for the free energy function, and consequently for the corresponding
stress tensor, ${\bf {Q}^{T}}{\bf {S}({\bf {C}},{\bf {\xi}})}{\bf {Q}}={\bf {S}({\bf {Q}^{T}{\bf {C}}{\bf {Q}},{Q}\star}{\xi})}$,
holds for $\forall{\bf {Q}}\in SO(3)$, which means that the function
is an \emph{isotropic} scalar-valued tensor function.
\item The relation $\psi({\bf {C}},{\bf {\xi}})=\psi({\bf {Q}^{T}{\bf {C}}{\bf {Q}},}{\xi})$
for the free energy function, and the corresponding stress tensor,
${\bf {Q}^{T}}{\bf {S}({\bf {C}},{\bf {\xi}})}{\bf {Q}}={\bf {S}({\bf {Q}^{T}{\bf {C}}{\bf {Q}},}{\xi})}$,
holds for $\forall{\bf {Q}}\in G_{M}$, which means that the function
is an \emph{anisotropic} scalar-valued tensor function.
\end{itemize}
On a more intuitive level, the difference between these two statements
is as follows: in the first one the deformation and the ``body
structure'' are both rotated with $\mathbf{Q}$, while in the second
statement only the deformation is rotated because the structural
tensor $\xi$ has the appropriate symmetry to rotations with
$\mathbf{Q} \in G_M$.


\section{Free energy function for cubic materials}
\label{sec:free_energy}

As we showed in the previous section the final aim in the proposal of
a model is to construct energy functions invariant under the
appropriate symmetry groups reflecting the underlying material
anisotropy. A direct way to do that is by means of functions dependent
on the invariants of the right Cauchy-Green tensor and structural
tensors, which ensures that the functions to be constructed are also
invariant under the proper symmetry group, retaining in this manner
the material symmetries of the body of interest. In particular we are
interested in proposing polynomial type energy functions.  In order to
keep the complexity to minimum we will make use of the minimal set of
independent invariants of the deformation and structural tensors.
This minimal set is called a polynomial basis and further details
about it can be found in \cite{Boehler:87}.

In the subsections to come we first present the structural tensor for
the particular case of a material with cubic symmetry and follow-up
with a description of the polynomial basis of invariants constructed
from this structural tensor and the right Cauchy-Green tensor
$\mathbf{C}$.

\subsection{Structural tensor for cubic anisotropy}

To determine the structural tensors corresponding to crystals with
cubic structure we follow Zheng (\cite{Zheng:93}), where the
structural tensors for all the different crystal classes are
developed. Zheng's paper is based on the available properties of
Kronecker products of orthogonal transformations which allow the
development of a simple method to determine the structural tensors
with respect to any given symmetry group. As it has been highlighted
in \cite{Zheng:93} there may exist many possible sets of the
structural tensors for a given symmetry group, so one goal set by the
author has been to find out the simplest irreducible representations;
in particular, it is shown that each of the anisotropic symmetry
groups can be characterized by a single structural tensor, and that is
the result we will make use of.

The crystal class corresponding to the Hexoctohedral cubic system is
characterized by the following generators of its finite symmetry group
\[
\mathbf{Q}_1(\frac{\pi}{2}),\mathbf{Q}_2(\frac{\pi}{2}),\mathbf{Q}_3(\frac{\pi}{2}), -\mathbf{1}
\]
where ${\bf {Q}}_{i}(\theta)$ refers to the rotation transformation
about the axis ${\bf {e}}_{i}$, corresponding to a positive oriented
orthonormal triad of vectors, through an angle $\theta$, and ${\bf {1}}$
is the second order identity tensor.

Considering these group generators and applying properties of Kronecker
products, the structural tensor for this particular cubic system is
determined to be (\cite{Zheng:93})
\begin{equation}
\mathbf{M} = \mathbf{e}^{(4)}_1 + \mathbf{e}^{(4)}_2 +
\mathbf{e}^{(4)}_3 \label{eq:structural-tensor}
\end{equation} 
where the notation $\mathbf{e}^{(4)}_i=\mathbf{e}_i \otimes
\mathbf{e}_i \otimes \mathbf{e}_i \otimes \mathbf{e}_i$ has been
introduced. The novelty of the approach suggested in this work is in
the use of this fourth order structural tensor $\mathbf{M}$ in
building up the energy function. All anisotropic structural tensors,
investigated in the literature so far, have been second order.

\subsection{Polynomial Basis}

An irreducible polynomial basis consists of a collection of members,
where none of them can be expressed as a polynomial function of the
others, i.e., they are independent scalars, and any other polynomial
invariant of the same tensors can be written as a polynomial function
of the basis members. The Hilbert theorem guarantees that a finite
integrity basis exists for any finite basis of tensors
(\cite{Weyl:46}).

Taking \cite{Scho-Neff:03} as a reference, we shall present an
analogous procedure for the construction of specific constitutive
equations based on functions whose arguments are the (joint)
invariants of the right Cauchy-Green tensor $\mathbf{C}$ and the
structural tensor $\mathbf{M}$. Next we present the integrity basis
invariants which will be the arguments of the constitutive functions
to be proposed.

The integrity basis consist of the traces of products of powers of the
argument tensors. They can be divided in two main groups: the
\emph{principal invariants}, which involve invariants of the
deformation tensor alone or the structural tensor alone, and the
so-called \emph{mixed invariants}, which consider joint invariants of
both tensors. In the following we present separately the different
kinds of invariants that can be formed from the right Cauchy-Green
tensor $\mathbf{C}$ and the structural tensor $\mathbf{M}$.
\begin{itemize}
\item Invariants of the right Cauchy-Green tensor alone

The principal invariants of the second order tensor ${\bf {C}}$,
denoted by $I_{k}=I_{k}({\bf {C}}),\,\, k=1,2,3$, are defined as the
coefficients of the characteristic polynomial
$f(\lambda)=\mathrm{det}\left[\lambda{\bf {1}}-{\bf {C}}\right]$ (see
Appendix \ref{sec:char_polynomial} for details). The explicit
expressions for the principal invariants of the second order tensor
$\mathbf{C}$ are
\[
\left\{ \begin{array}{l} I_{1}\equiv
\mathrm{tr}\left(\mathbf{C}\right)\\ I_{2}\equiv
\mathrm{tr}\left(\mathrm{cof}\mathbf{C}\right)\\ I_{3}\equiv
\mathrm{det}\left(\mathbf{C}\right)\end{array}\right.\] which can be
expressed in terms of the \emph{basic invariants} $J_{i},\,\,
i=1,2,3$, defined as the traces of powers of ${\bf {C}}$ :
\[ \left\{ \begin{array}{l} J_{1}\equiv
\mathrm{tr}\left({\bf {C}}\right)={\bf {1}}:{\bf {C}}\\ J_{2}\equiv
\mathrm{tr}\left({\bf {C}^{2}}\right)={\bf {1}}:{\bf {C}}^{2}\\
J_{3}\equiv \mathrm{tr}\left({\bf {C}^{3}}\right)={1}:{\bf
{C}}^{3}\end{array}\right.\]
\item Mixed invariants of the right Cauchy-Green and the structural
tensors

In the case of several tensor variables, we use the term mixed invariant,
even though the term simultaneous invariant can also be found in the
literature (\cite{Truesdell:65}).

We will follow Betten (\cite{Betten:81}) to determine the scalar
invariants of the tensors ${\bf {C}}$ and ${\bf {M}}$. To construct a
set of mixed invariants of the second-order tensor ${\bf {C}}$ and the
fourth-order structural tensor ${\bf {M}}$ we consider a theorem
presented in the reference \cite{Betten:81} and its generalization for
fourth-order tensors using the Hamilton-Cayley theorem, which means
that powers of tensor ${\bf {M}}^{n}$ and higher can be expressed in
terms of ${\bf \mathbf{1},{\bf {M}}},$${\bf {M}}^{2},...,{\bf
{M}^{n-1}}$, where $n$ represents the vector space dimension of ${\bf
{C}}$ (for a symmetric second-order tensor $n=6$).  The additional
mixed invariants are

\begin{equation}
\left\{ \begin{array}{l}
I_{4}^{k}\equiv{\bf {C}}:{\bf {M}}^{k}:{\bf {C}}\\
I_{5}^{k}\equiv{\bf {C}}:{\bf {M}}^{k}:{\bf {C}}^{2}\\
I_{6}^{k}\equiv{\bf {C}}^{2}:{\bf {M}}^{k}:{\bf {C}}^{2},
\end{array}\right.\label{eq:invar-js}\end{equation}
where $k=1,2,3,4,5$.  As shown in \cite{Betten:81}, the proof of the
theorem relies on the assumption that $\mathbf{M}$ satisfies the
symmetry conditions
\begin{equation}
M_{IJKL}=M_{JIKL}=M_{IJLK}=M_{KLIJ}.\label{eq:M-symmetries}
\end{equation}
Clearly these conditions are fulfilled by the structural tensor
$\mathbf{M} = \mathbf{e}^{(4)}_1+ \mathbf{e}^{(4)}_2 +
\mathbf{e}^{(4)}_3$.

In addition to the invariants (\ref{eq:invar-js}), \cite{Betten:81}
 proved that the following expressions are also invariant

\begin{equation}
\left\{ \begin{array}{l} \bar{I}^k_M \equiv \mathbf{1}:
  \mathbf{M}^{k}:\mathbf{C} \\ \bar{\bar{I}}^k_M \equiv \mathbf{1}:
  \mathbf{M}^k: \mathbf{C}^2 . \end{array}\right.
\label{eq:bar-inv}
\end{equation} 
We will skip writing the superscript $k$ in (\ref{eq:invar-js}) and
(\ref{eq:bar-inv}) when $k=1$.

Additionally we will also need the invariant
\begin{displaymath}
\hat{I}_M \equiv \mathbf{1}:\mathbf{M}:\mathrm{adj}(\mathbf{C}).
\end{displaymath}

\item Invariants of the fourth-order structural tensor alone

The only remaining basic invariant of the single tensor ${\bf {M}}$,
formed as $\mathrm{tr}({\bf {M}})$, is constant, and therefore it is
not useful for the construction of strain energy functions.
\end{itemize}

\subsection{Polyconvexity condition\label{sub:Polyconvexity-condition}}

In this section we briefly describe sufficient (but not necessary)
free energy function conditions which guarantee the existence of
minimizers of some variational principles for finite strains. As
already mentioned, polyconvexity is the property of interest to us.

Local existence and uniqueness theorems in nonlinear elastostatics and
elastodynamics are based on strong ellipticity. The ellipticity
condition states that the elastic free energy $\psi({\bf {F}})$ leads
to an elliptic system if and only if the Legendre-Hadamard condition
\[
\forall{\bf {F}},\forall\xi,\eta\in\Re^{3}:\quad D_{{\bf {F}}}^{2}\psi({\bf {F}})(\xi\otimes\eta,\xi\otimes\eta)\geq0\]
holds.

The early global existence theory for elastostatics was based on
convexity of the free energy function. However, that condition, as
shown in \cite{Ball:77}, is unreasonable from a physical point of
view. Using the notion of quasiconvexity due to Morrey
(\cite{Morrey:52}), Ball (\cite{Ball:77}) proved global existence
theorems for elastostatics.  In particular, it was proven that
quasiconvexity implies the existence of minimizers of some variational
principles in finite elasticity.  The quasiconvexity condition reads
\[
\forall{\bf {F}},\forall\omega\in C_{0}^{\infty}(B)\qquad\psi({\bf
{F}})\left|B\right|={\displaystyle \int_{B}\psi({\bf
{F}})}dV\leq{\displaystyle \int_{B}\psi({\bf {F}}+\nabla\omega)}dV
\]
Unfortunately this integral inequality is complicated to handle. A
concept of greater practical importance is that of polyconvexity in
the sense of Morrey (\cite{Ball:77}). Following Marsden
(\cite{MARSDEN:83}), we say that the energy function $\psi$ is
polyconvex if and only if there exits a function $\varphi$ with
arguments ${\bf {F}}$, $\mathrm{adj}({\bf
{F}})=\mathrm{det}(\mathbf{F})\mathbf{F}^{-1}$ and $\mathrm{det}({\bf
{F}})$ such that
\begin{equation}
\psi({\bf {F}})=\varphi({\bf {F}},\mathrm{adj}({\bf
  {F}}),\mathrm{det}({\bf {F}}))
\label{eq:polyconvexity}
\end{equation}
and $\varphi$ is convex function.

As an illustrative example we present the case of $\psi({\bf
{F}})=f(\mathrm{det}{\bf {F}})$, for a convex function $f$. This
function is not convex taken as a function of ${\bf {F}}$ (because the
range of definition of ${\bf {F}}$ is not convex), however, it
fulfills the polyconvex condition, since the condition of
polyconvexity requires to take $\mathrm{det}({\bf {F}})$ as the
independent variable and, by hypothesis, the function $f$ is convex in
that variable.  The polyconvexity condition has additive nature, i.e.,
if the functions $\psi_{i},\, i=1,2,3$ are all convex in their
respective arguments then the function $\psi({\bf {F}})=\psi_{1}({\bf
{F}})+\psi_{2}(\mathrm{adj}({\bf {F}}))+\psi_{3}(\mathrm{det}({\bf
{F}}))$ is polyconvex. This property turns out to be very useful when
proposing models because it permits to construct energy functions out
of simpler ones.

Due to the material indifference condition, the dependency on ${\bf
{F}}$ of the energy function can be completely replaced by dependency
on ${\bf {C}}$, but the polyconvexity condition does not translate to
functions of $\mathbf{C}$ in a simple manner.

Finally, we summarize the implication chain relating all the previous
concepts
\[
\textrm{convexity}\Rightarrow\textrm{polyconvexity}\Rightarrow\textrm{quasiconvexity}\Rightarrow\textrm{ellipticity}
\]
None of the opposite implications is true as counter-examples have
been found for all of them (\cite{MARSDEN:83}).

\subsection{Isotropic free energy terms}


For completeness, here we present two statements about polyconvexity of
some simple isotropic functions.  The interested reader should consult
\cite{Scho-Neff:03} about further details on polyconvexity of
various isotropic functions.

\begin{description}
\item [Statement.]The polynomial function

\begin{equation}
{\bf {F}}\mapsto\gamma(\mathrm{tr}(\mathbf{F}^T \mathbf{F}))^{k}=\gamma
I_{1}^{k}\,\,\,\,\,\,\,\textrm{with$\,k\geq 1 \textrm{ and}\,\,\gamma>0$}
\label{eq:first-iso}
\end{equation}
 is polyconvex.
\item [Proof.] The function $(\mathrm{tr}(\mathbf{F}^T
\mathbf{F}))^k=\left \Vert \mathbf{F} \right \Vert ^{2k}$ can be
considered to be a function of $\mathbf{F}$ only and therefore,
referring to the results in the previous section, it is enough to prove
convexity relative to the argument $\mathbf{F}$.  As described in
\cite{Scho-Neff:03} the one possible approach to show convexity is to
check the positivity of the second G\^ateaux derivative:

\[
<D\left(\mathbf{F}^T \mathbf{F} \right)^{k},{\bf {H}}>={\displaystyle
  \frac{d}{d\epsilon}}\left.\left[\left({\bf {F}+{\bf
      \epsilon{H}}}\right)^{T}\left({\bf {F}+{\bf
      \epsilon{H}}}\right)\right]^{k}\right|_{\epsilon=0}=2k\left\Vert
{\bf {F}}\right\Vert ^{2k-2}<{\bf {F}},{\bf {H}}>,
\]
and from here the second derivative yields

\[
\begin{array}{l}
<D^{2}\left( \mathbf{F}^T \mathbf{F}
\right)^{k},\left({H},{H}\right)>={\displaystyle
  \frac{d}{d\epsilon}}\left.<D\left(\mathbf{F}^T \mathbf{F} \right)^{k},{\bf {H}}>\right|_{\epsilon=0}\\
\\\qquad=2k\left(\left\Vert {\bf {F}}\right\Vert ^{2k-2}<{\bf
  {H}},{\bf {H}}>+(2k-2)\left\Vert {\bf {F}}\right\Vert ^{2k-4}<{\bf
  {F}},{\bf {H}}>^{2}\right)\ge0 .\end{array}
\]
The desired result is established by noting that the constant $\gamma$
is positive and does not modify the signs of the derivatives.$\Box$
\end{description}

\begin{description}
\item [Statement.]The functions
\[
{\bf {F}}\mapsto\gamma(\mathrm{det}(\mathbf{F}^T \mathbf{F}))^{k}=\gamma
I_{3}^{k},\,\,\,\,\,\,\,\textrm{with $k\ge 1$, $\,\,\gamma>0$ and}\]
\[
\mathbf{F}\mapsto -\beta \log(\det(\mathbf{F}^T \mathbf{F})) = -\beta \log(I_3),\,\,\,\,\,\,\,
\textrm{with $\beta>0$}\,\,\,
\]
 are polyconvex.
\item [Proof.] After noting that $I_3=(\det(\mathbf{F}))^2$, it is
sufficient to show convexity relative to $\mathrm{det}(\mathbf{F})$.
The convexity is established by checking the non-negativity of the
second derivatives of $x^{2k}$ and $-2\log(x)$ which is a trivial
exercise. $\Box$
\end{description}

\subsection{Anisotropic free energy terms}

Next we analyze the polyconvexity of some terms dependent on the
structural tensor $\mathbf{M}$.

\begin{description}
\item [Statement.]The polynomial function 
\[
{\bf {F}}\mapsto\gamma\left(\mathrm{tr}\left(\mathbf{F}^T \mathbf{F}
 \mathbf{M} \mathbf{F}^T \mathbf{F} \right)\right)^{k}=\gamma I_{4}^{k}\textrm{ \,\,\,\,\,\,\, with $k\geq1$ and$\,\,\,\,\gamma>0$}\]
 is polyconvex.
\item [Proof.] Mimicking the approach used for $I_1^k$, we proceed by
showing that $I_4^k$ is a convex function of $\mathbf{F}$.  Given that
\[
\left(\mathrm{tr}\left(\mathbf{F}^T \mathbf{F} \mathbf{M} \mathbf{F}^T
\mathbf{F} \right)\right)^{k}=\left((\mathbf{F}^T \mathbf{F}
):\mathbf{M}:(\mathbf{F}^T \mathbf{F})\right)^{k}\]
we can obtain the first and second G\^ateaux derivatives of $I_4^k$:

\begin{displaymath}
\begin{array}{l}
<D\left((\mathbf{F}^T \mathbf{F}):\mathbf{M}:(\mathbf{F}^T \mathbf{F}
)\right)^{k},{\bf {H}}>=\\ ={\displaystyle
\frac{d}{d\epsilon}}\left.\left[\left({\bf {F}+{\bf
\epsilon{H}}}\right)^{T}\left({\bf {F}+{\bf \epsilon{H}}}\right):{\bf
{M}}:\left({\bf {F}+{\bf \epsilon{H}}}\right)^{T}\left({\bf {F}+{\bf
\epsilon{H}}}\right)\right]^{k}\right|_{\epsilon=0}= \\
\begin{array}{l}
=2k\left((\mathbf{F}^T \mathbf{F} ):\mathbf{M}:(\mathbf{F}^T
    \mathbf{F} )\right)^{k-1}\left(\mathbf{F}^T \mathbf{H}:{\bf
    {M}}:\mathbf{F}^{T}\mathbf{F}+ \mathbf{H}^T \mathbf{F} :{\bf {M}}:\mathbf{F}^{T} \mathbf{F}\right)
\end{array}\end{array}
\end{displaymath}

\begin{equation}
\begin{array}{l}
<D^{2}\left((\mathbf{F}^T \mathbf{F}):\mathbf{M}:(\mathbf{F}^T
\mathbf{F} )\right)^{k},\left({H},{H}\right)>={\displaystyle
  \frac{d}{d\epsilon}}\left.<D\left((\mathbf{F}^T
\mathbf{F}):\mathbf{M}:(\mathbf{F}^T \mathbf{F})\right)^{k},{\bf {H}}>\right|_{\epsilon=0}\\
\begin{array}{l}
=4k(k-1)\left((\mathbf{F}^T \mathbf{F}):\mathbf{M}:(\mathbf{F}^T
\mathbf{F} )\right)^{k-2}\left[\mathbf{F}^T \mathbf{H}:{\bf
    {M}}:\mathbf{F}^{T}\mathbf{F}+ \mathbf{H}^T \mathbf{F} :{\bf {M}}:\mathbf{F}^{T}\mathbf{F}\right]^2\\
\begin{array}{l}
+2k\left((\mathbf{F}^T \mathbf{F} ):\mathbf{M}:(\mathbf{F}^T \mathbf{F})\right)^{k-1}\times
    [2\mathbf{H}^T\mathbf{H}:\mathbf{M}:\mathbf{F}^T\mathbf{F}
    +\mathbf{F}^T\mathbf{H}:\mathbf{M}:\mathbf{H}^T\mathbf{F}+\\ + \mathbf{F}^T\mathbf{H}:\mathbf{M}:\mathbf{F}^T\mathbf{H}
+\mathbf{H}^T\mathbf{F}:\mathbf{M}:\mathbf{H}^T\mathbf{F}+\mathbf{H}^T\mathbf{F}:\mathbf{M}:\mathbf{F}^T\mathbf{H}]=\\
  \end{array}\end{array}\\
\begin{array}{l}
\begin{array}{l}
\begin{array}{l}
=16k(k-1)\left((\mathbf{F}^T \mathbf{F}):\mathbf{M}:(\mathbf{F}^T
    \mathbf{F} )\right)^{k-2}\left[\mathbf{F}^T \mathbf{H}:{\bf
    {M}}:\mathbf{F}^{T}\mathbf{F}\right]^{2}+
\end{array}\\
\begin{array}{l}
+4k\left((\mathbf{F}^T \mathbf{F}):\mathbf{M}:(\mathbf{F}^T \mathbf{F}
    )\right)^{k-1}\left[\mathbf{H}^T\mathbf{H}:{\bf
    {M}}:\mathbf{F}^{T}\mathbf{F}+2 \mathbf{H}^T \mathbf{F} :{\bf
    {M}}:\mathbf{H}^T \mathbf{F}\right] , \end{array}\end{array}
    \end{array}\end{array} \label{secondder_i4}
\end{equation}
where the last equality used the symmetry properties of the structural
tensor $\mathbf{M}$, more specifically relations (\ref{eq:A1}),
(\ref{A2}) and (\ref{eq:A3}). To complete the proof we separately
analyze each term of the second derivative and show its
non-negativity.

The non-negativity of $I_4$ follows from:
\begin{eqnarray*}
I_4 & = & (\mathbf{F}^T \mathbf{F}):\mathbf{M}:(\mathbf{F}^T \mathbf{F}) = 
\sum_{i=1}^3 (\mathbf{F}^T \mathbf{F}):\mathbf{e}_i^2 \otimes \mathbf{e}_i^2 :
(\mathbf{F}^T \mathbf{F}) = \\
& = & \sum_{i=1}^3 [\mathbf{F}^T \mathbf{F} : \mathbf{e}_i \otimes \mathbf{e}_i]^2
\ge 0,
\end{eqnarray*}
where property (\ref{eq:ABBC}) has been used.  In a similar manner,
making use of both (\ref{eq:ABBC}) and (\ref{eq:eeATA}), we show that
the other terms participating in the expression for the second
derivative are also non-negative:
\begin{eqnarray*}
\mathbf{H}^T\mathbf{H}:\mathbf{M}:\mathbf{F}^T\mathbf{F} & = &
\sum_{i=1}^3 (\mathbf{H}^T\mathbf{H} : \mathbf{e}_i^2)(\mathbf{e}_i^2: \mathbf{F}^T
\mathbf{F} ) = \sum_{i=1}^3 (\mathbf{He}_i)^2 (\mathbf{Fe}_i)^2 \ge 0 ,\\
\mathbf{H}^T\mathbf{F}:\mathbf{M}:\mathbf{H}^T\mathbf{F} & = &
\sum_{i=1}^3 (\mathbf{H}^T \mathbf{F} : \mathbf{e}_i \otimes \mathbf{e}_i)^2 
\ge 0 .
\end{eqnarray*}
The non-negativity of all terms in the expression for the second
derivative together with the nonnegativity of $\gamma$ implies the
desired result. $\Box$

\end{description}

Next, analogously to \cite{Scho-Neff:03}, we prove the polyconvexity of
$\frac{I_4}{I_3^{2/3}}$ . This is the isochoric term corresponding to
$I_4$.

\begin{description}
\item [Statement.]The function
\[ 
\mathbf{F} \mapsto \gamma \frac{\mathrm{tr}\left(\mathbf{F}^T
    \mathbf{FMF}^T\mathbf{F} \right)}{\mathrm{det}(\mathbf{F})^{4/3}}
    = \gamma \frac{I_{4}}{I_{3}^{2/3}}\;\;\;\mathrm{with}\;\; \gamma > 0
\]
is polyconvex.

\item [Proof.] Let $\phi(x,y)=\frac{x^4}{y^{4/3}}$ and 
\begin{displaymath}
  \psi_{\eta}({\mathbf F},\zeta)=\phi(\left\Vert{\mathbf F
  \eta}\right\Vert,\zeta)=\frac{\left\Vert{\mathbf F \eta}\right\Vert^4}{\zeta^{4/3}},
\end{displaymath}
where ${\mathbf{\eta}}$ is an arbitrary vector. We will establish that
$\psi_\eta$ is a convex function, when considered as a function of both
arguments simultaneously. Condition (\ref{eq:neff-lemma1}) is
satisfied for $p=4$ and $\alpha=4/3$, therefore the function
$\phi(x,y)$ is convex. The convexity of $\psi_{\eta}({\mathbf
F},\zeta)$ follows from the following sequence of inequalities:
\begin{displaymath}
  \psi_{\eta}(\lambda{\mathbf F_1}+(1-\lambda){\mathbf F_2},\lambda
  \zeta_1+(1-\lambda)\zeta_2)=\frac{\left\Vert\lambda{\mathbf
  F_1\eta}+(1-\lambda){\mathbf F_2\eta}\right\Vert^4}{(\lambda
  \zeta_1+(1-\lambda)\zeta_2)^{4/3}}\le
\end{displaymath}
\begin{displaymath}
  \frac{(\lambda\left\Vert{\mathbf
  F_1\eta}\right\Vert+(1-\lambda)\left\Vert{\mathbf F_2\eta}\right\Vert)^4}{(\lambda
  \zeta_1+(1-\lambda)\zeta_2)^{4/3}}=\phi(\lambda\left\Vert{\mathbf
  F_1\eta}\right\Vert+(1-\lambda)\left\Vert{\mathbf F_2\eta}\right\Vert,\lambda
  \zeta_1+(1-\lambda)\zeta_2)\le
\end{displaymath}
\begin{displaymath}
  \lambda\phi(\left\Vert{\mathbf F_1
  \eta}\right\Vert,\zeta_1)+(1-\lambda)\phi(\left\Vert{\mathbf F_2
  \eta}\right\Vert,\zeta_2)=\lambda\psi_{\eta}({\mathbf
  F_1},\zeta_1)+(1-\lambda)\psi_{\eta}({\mathbf F_2}),
\end{displaymath}
where the triangular inequality and the convexity of $\phi(x,y)$ have
been used.  The required result can be obtained directly from:
\begin{displaymath}
\frac{I_4}{I_3^{2/3}}=\frac{{\mathbf C}:({\mathbf
  e_1}^4+{\mathbf e_2}^4+{\mathbf e_3}^4):{\mathbf C}}{({\mathrm{
  det}}({\mathbf F}))^{4/3}}=\frac{\left\Vert{\mathbf{F
  e_1}}\right\Vert^4+\left\Vert{ \mathbf{F
  e_2}}\right\Vert^4+ \left\Vert{\mathbf{F e_3}}\right\Vert^4}{({\mathrm{det}}({\mathbf F}))^{4/3}}
\end{displaymath}
\begin{displaymath}
= \psi_{\mathbf{e}_1}({\mathbf F},{\mathrm{det}}({\mathbf
  F}))+ \psi_{\mathbf{e}_2}({\mathbf F},{\mathrm{det}}({\mathbf F}))+ \psi_{\mathbf{e}_3}({\mathbf F},{\mathrm{det}}({\mathbf F})).
\end{displaymath}
The positive coefficient $\gamma$ does not influence the conclusion.  The same
proof can be applied to $\gamma\frac{I_4}{I_3^{\alpha/2}}$ as long as
$0\le \alpha \le 3$.  $\Box$
\end{description}

As pointed out in \cite{Scho-Neff:03} the apparent symmetry between
$\mathbf{F}$ and $\mathrm{adj}(\mathbf{F})$ in the definition
(\ref{eq:polyconvexity}) suggests that new polyconvex functions can be
obtained by switching the deformation gradient tensor $\mathbf{F}$
with its adjoint tensor $\mathrm{adj}(\mathbf{F})$.  The reader must
be warned that replacing $\mathbf{C}$ with $\mathrm{adj}(\mathbf{C})$
is not equivalent to replacing $\mathbf{F}$ with
$\mathrm{adj}(\mathbf{F})$ because
\[
\mathrm{adj}(\mathbf{C})=\mathrm{det}(\mathbf{C})\mathbf{C}^{-1}=
\mathrm{det}(\mathbf{F})^2 (\mathbf{F}^T\mathbf{F})^{-1} = 
\mathrm{det}(\mathbf{F})^2 \mathbf{F}^{-1}\mathbf{F}^{-T} = 
\mathrm{adj}(\mathbf{F}) \mathrm{adj}(\mathbf{F})^T.
\]
The difference comes from the position of the transpose symbol: in
$\mathbf{C}=\mathbf{F}^T \mathbf{F}$ the first matrix is transposed,
while in $\mathrm{adj} (\mathbf{C}) = \mathrm{adj}(\mathbf{F})
\mathrm{adj} (\mathbf{F})^T$ it is the second one, but the proofs
already presented in this section are independent of the position of
the transpose symbol and we will exploit this to verify the following


\begin{description}
\item[Statement.] Let $K_3=\mathrm{adj}(\mathbf{C}) : \mathbf{M} :
\mathrm{adj}(\mathbf{C})$, then the functions
\begin{eqnarray*}
\mathbf{F} & \mapsto & \gamma \left( \mathrm{adj}(\mathbf{F}^T \mathbf{F})
 :\mathbf{M} : \mathrm{adj}(\mathbf{F}^T \mathbf{F}) \right)^k =
 \gamma K_3^k\;\;\;
\mathrm{with}\;\; k\ge 1 \;\; \mathrm{and} \;\; \gamma > 0,\\
\mathbf{F} & \mapsto & \gamma \frac{ \mathrm{adj}(\mathbf{F}^T \mathbf{F})
 :\mathbf{M} : \mathrm{adj}(\mathbf{F}^T \mathbf{F}) }{I_3^{4/3}} =
 \gamma \frac{K_3}{I_3^{4/3}} \;\;\; \mathrm{with} \;\; \gamma > 0
\end{eqnarray*}
are polyconvex.
\item[Proof.] If $K_3$ is a scalar invariant of $\mathbf{C}$ and
$\mathbf{M}$ then the two functions will be polyconvex because the
proofs of the previous two statements are independent of the position
of the transpose symbol and will not be repeated here (for the second
function, $\alpha=8/3$ is in the range of values which preserve the
validity of the previous statement).  The function
$\frac{K_3}{I_3^{4/3}}$ is the proper isochoric variant of $K_3$.

We only need to prove that $K_3$ is a scalar invariant of $\mathbf{C}$
and $\mathbf{M}$.  We will prove on the way that $K_1=\mathrm{adj}(
\mathbf{C}) :\mathbf{M} : \mathbf{C}$, $K_2=\mathrm{adj}( \mathbf{C})
:\mathbf{M} : \mathbf{C}^2$ and $\hat{I}_M=\mathbf{1} : \mathbf{M} :
\mathrm{adj} (\mathbf{C} )$ are also scalar invariants. If the
characteristic polynomial (\ref{eq:cayley-hamilton}) is multiplied as
a double scalar product by $\mathbf{C}^{-1}\mathbf{M} : \mathbf{C}$,
then the following equation for $K_1$ is obtained
\begin{displaymath}
K_1= I_5-I_1 I_4 +I_2 \bar{I}_M.
\end{displaymath}
If $K_1$ can be expressed as a combination of scalar invariants of
$\mathbf{M}$ and $\mathbf{C}$ , then $K_1$ is a scalar invariant
itself.

Proceeding in the same manner for $K_2$, one obtains after
multiplication by $\mathbf{C}^{-1}\mathbf{M} : \mathbf{C}^2$
\begin{displaymath}
K_2= I_6-I_1 I_5 +I_2 \bar{\bar{I}}_M.
\end{displaymath}

For $K_3$ the multiplication is by $\mathbf{C}^{-1}\mathbf{M} :
\mathrm{adj} (\mathbf{C})$ and the result is
\begin{displaymath}
K_3= K_2-I_1 K_1 +I_2 \hat{I}_M.
\end{displaymath}
Given that $K_1$ and $K_2$ are scalar invariants, for $K_3$ to be a
scalar invariant one needs to show that $\hat{I}_M$ is a scalar
invariant.  This can be accomplished by multiplication by
$\mathbf{C}^{-1}\mathbf{M} : \mathbf{1}$ with the result being
\begin{displaymath}
\hat{I}_M= \bar{\bar{I}}_M-I_1 \bar{I}_M +I_2.
\end{displaymath}
In fact, for our specific tensor $\mathbf{M} =\mathbf{e}_1^4
+\mathbf{e}_2^4 +\mathbf{e}_3^4$ the invariant $\hat{I}_M$ is equal to
the invariant $I_2$. $\Box$
\end{description}

\section{Model for polyconvex free energy function with cubic
  anisotropy}
\label{sec:model}

Having presented proofs of polyconvexity for some strain energy
functions, we proceed to propose a global model based on these
functions for the case of materials with cubic anisotropy. The
mathematical sanity of the model is guaranteed in advance by means of
the polyconvexity of the proposed strain energy function.
Consequently, this allows the application of the theorems concerning
the existence of minimizing sequences discussed in Section
\ref{sub:Polyconvexity-condition}.

The proposed model is a linear combination of the energy functions
proved above to be polyconvex. The general form of the free energy
function reads

\begin{eqnarray}
\psi=\underbrace{\alpha (-\log(I_3))}+\underbrace{\beta
  I_{3}}+\underbrace{\gamma I_{4}} +\underbrace{\delta I_1}
\label{Model}\\
\psi_1 \;\;\;\;\;\;\;\;\;\;\;\; \psi_2 \;\;\;\;\; \psi_3
\;\;\;\;\;\; \psi_4 \; \nonumber
\end{eqnarray}
where the last three terms have been selected to be as simple as
possible, even though polyconvexity was shown for more general cases
with exponents larger than 1.

For completeness we briefly present the relationship between stresses
and the free energy function based on the expression
(\ref{eq:s-derivpsi}).  Making use of the additively decoupled
formulation of the polyconvex function (\ref{Model}) we expand the
stress tensor in the following general manner

\begin{equation}
{\bf {S}}=2\frac{\partial\psi}{\partial{\bf {C}}}=2{\displaystyle
\sum_{i=1}^{4}\sum_{j=1}^6}\frac{\partial\psi_{i}}{\partial
I_{j}}\frac{\partial I_{j}}{\partial{\bf
{C}}}\label{eq:stress-divi}\end{equation}

From (\ref{eq:stress-divi}), following a procedure analogous to
(\ref{eq:s-intermedi}), we can deduce the formal expression for the
tangent matrix:

\begin{displaymath}
{\bf {\mathbb{C}}} = \frac{\partial{\bf {S}}}{\partial{\bf
    {F}}}=\frac{\partial{\bf {C}}}{\partial{\bf
    {F}}}\frac{\partial{\bf {S}}}{\partial{\bf
    {C}}}=2\frac{\partial{\bf {S}}}{\partial{\bf
    {C}}}=4\frac{\partial{\bf ^{2}\psi}}{\partial{\bf
    {C}}^{2}}=
\end{displaymath}
\begin{equation}
  = 4{\displaystyle \sum_{i=1}^{4}\sum_{j=1}^6}\left[\frac{\partial
    \psi_{i}}{\partial I_{j}}\frac{\partial^2 I_{j}}{\partial{\bf
    {C}}^2}+\sum_{k=1}^6 \frac{\partial^2\psi_{i}}{\partial I_{j}
    \partial I_k}\frac{\partial I_{j}}{\partial{\bf
    {C}}}\frac{\partial I_{k}}{\partial{\bf
    {C}}}\right]\label{eq:tangent-divi}\end{equation}

Expressions for each individual term corresponding to the first and
second derivatives of the invariants can be found in Appendix
\ref{sub:Invariants-derivatives}.

\section{Adittional Conditions}
\label{sec:conditions}

In order to make a connection between the model and the physical data some
conditions will be imposed on the model.  These conditions will help
determine uniquely the values of the arbitrary constants accompanying
the individual functions appearing in the proposed model.

The conditions discussed here refer to the comparison between the
model response and the values of the parameters characterizing the
physical behavior of a material in a natural state. The natural state
chosen in this model corresponds to unstressed, undeformed
configuration, i.e. ${\bf {C}}={\bf {1}}$.

With the help of (\ref{eq:stress-divi}) and (\ref{eq:tangent-divi})
the following two conditions can be formulated:


\begin{itemize}
\item \textbf{Stress free reference configuration}  

This condition states that stresses must be zero when the deformation
gradient becomes unity.  Physically, this is equivalent to not having
any remanent tensions when the material is totally unloaded.
Mathematically, the stress free reference configuration means ${\bf
{S}}({\bf {1}})={\bf {0}}$, or upon substitution of the numerical
values into (\ref{eq:stress-divi})
\begin{equation}
-\alpha+\beta+2\gamma+\delta=0 \label{eq:zerostress} .
\end{equation}

\item \textbf{Tangent matrix at reference configuration}

The operation to be performed is the identification between the
tangent matrix (\ref{eq:tangent-divi}), particularized at the origin,
${\bf {C}}={\bf {1}}$, with the physical values corresponding to the
classical elastic moduli matrix of a cubic material.  There is an
implicit assumption that the values of the elastic constants remain
steady (time independent) for different values of the tensor
$\mathbf{C}$; albeit not being a completely realistic assumption, it
can be considered a well-posed first approximation in the development
of the model.  Thus, we have the identification
\[
{\mathbb{C}}_{0}\equiv\left.2\partial_{{\bf {C}}}{\bf
{S}}\right|_{{\bf {C}}={\bf {1}}}=\left.4\partial_{{\bf
{C}}}^{2}\psi\right|_{{\bf {C}}={\bf {1}}}\] where ${\mathbb{C}}_{0}$
represents the tangent matrix at the origin.  Substitution of the
numerical values for the derivatives gives the following three
equations:
\begin{eqnarray}
\alpha + 2\gamma & = & \mathcal{C}_{11}\\
\beta & = & \mathcal{C}_{12}\\
\alpha-\beta & = & 2\mathcal{C}_{44}, \label{eq:c44}
\end{eqnarray}
where $\mathcal{C}_{11}$, $\mathcal{C}_{12}$ and $\mathcal{C}_{44}$
are the standard elasticity constants in Voigt notation.
\end{itemize}

The solution of the system of equations
(\ref{eq:zerostress}-\ref{eq:c44}) is:
\begin{eqnarray*}
\alpha & = & \mathcal{C}_{12}+2\mathcal{C}_{44}\\
\beta & = & \mathcal{C}_{12} \\
\gamma & = &
\frac{1}{2}(\mathcal{C}_{11}-\mathcal{C}_{12}-2\mathcal{C}_{44})\\
\delta & = & -\mathcal{C}_{11}+\mathcal{C}_{12} + 4\mathcal{C}_{44}.
\end{eqnarray*}
The nonnegativity of the elastic constants implies nonnegativity of
$\alpha$ and $\beta$.  For $\gamma$ and $\delta$ to be nonnegative the
following condition must be satisfied:
\begin{equation}
\frac{1}{2}\le \frac{2
  \mathcal{C}_{44}}{\mathcal{C}_{11}-\mathcal{C}_{12}}\le 1.
\label{eq:condition}
\end{equation}
Conversely, if the inequalities (\ref{eq:condition}) are satisfied then
$\alpha$, $\beta$, $\gamma$ and $\delta$ are non-negative.  The ratio
$A=\frac{2 \mathcal{C}_{44}} {\mathcal{C}_{11}-\mathcal{C}_{12}}$ is
called the anisotropy ratio (see \cite{Hirth:82}) and its emergence in
condition (\ref{eq:condition}) suggests that the relevant measures of
anisotropy are an integral part of the model itself.  The data
available in Hirth (\cite{Hirth:82}) shows that only five
transitionary metals which are next to each other in the periodic
table satisfy both inequalities: Cr, Mo, W, V, Nb.  All five of them
have body-centered cubic crystal structure.  Some compound solids with
cubic structure like AgBr and NaCl also satisfy the inequalities
(\ref{eq:condition}).

The polyconvexity properties of $I_5$ and $I_6$ are currently unknown
to the authors, but the addition of terms proportional to $I_5$ or
$I_6$ in the strain energy function will not modify the condition
(\ref{eq:condition}) and consequently not enlarge the set of materials
to which the model can be applied.

\section{Study in 1D}
\label{sec:1dstudy}

\begin{figure}
\includegraphics[width=2.75in]{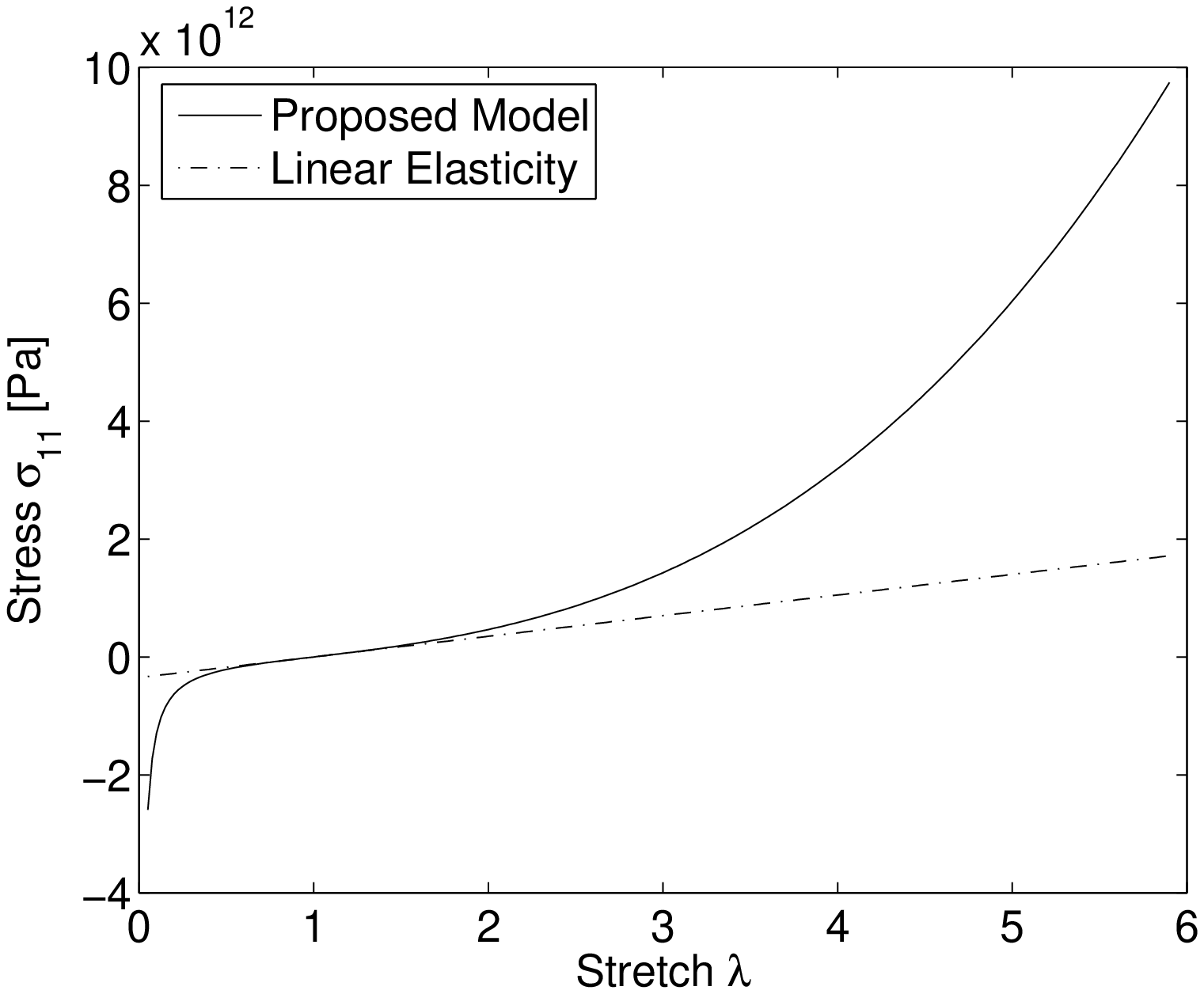}
\includegraphics[width=2.75in]{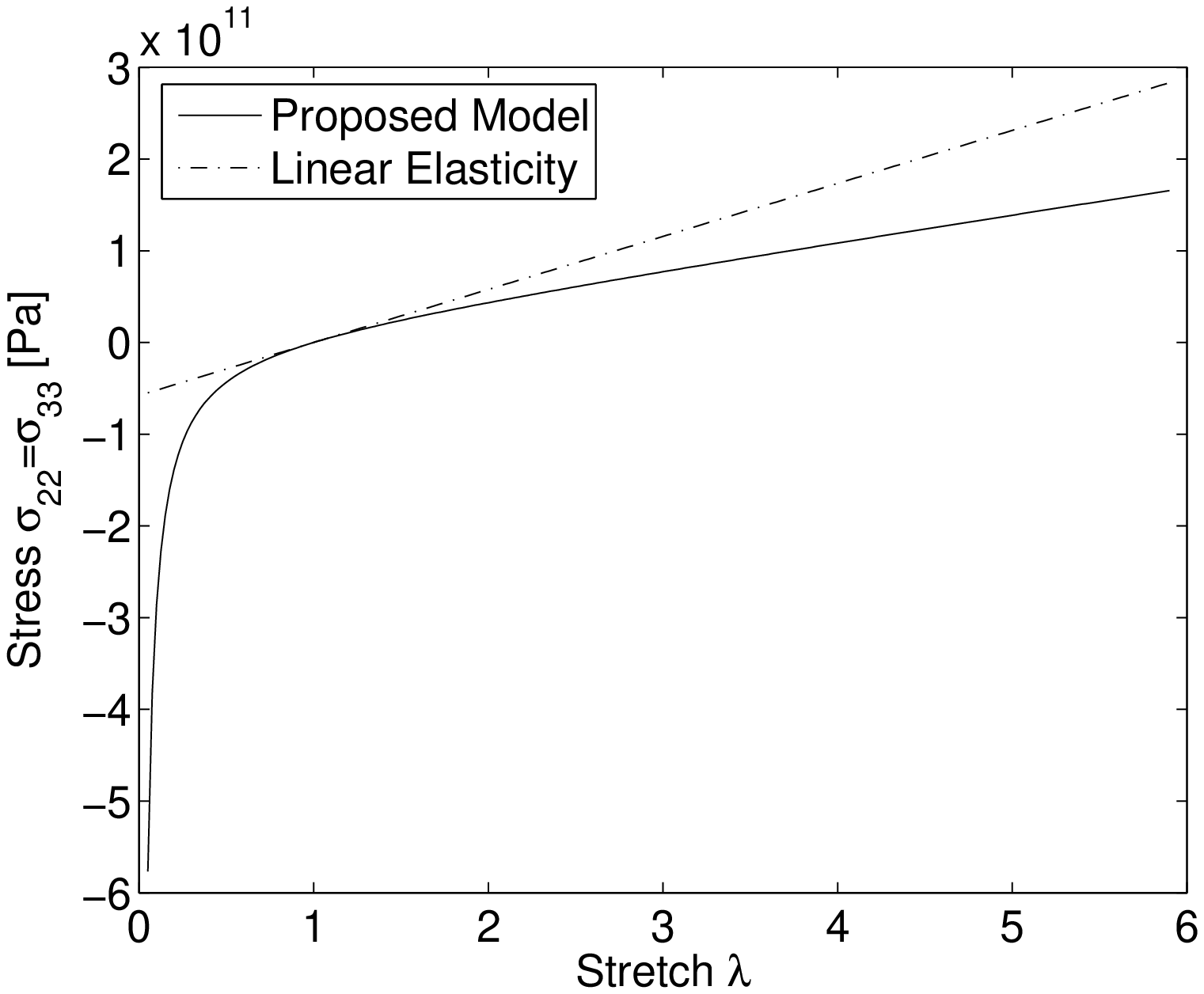}
\caption{Model response to simple stretch}
\label{fig:modelcmp}
\end{figure}

\begin{figure}
\includegraphics[width=2.75in]{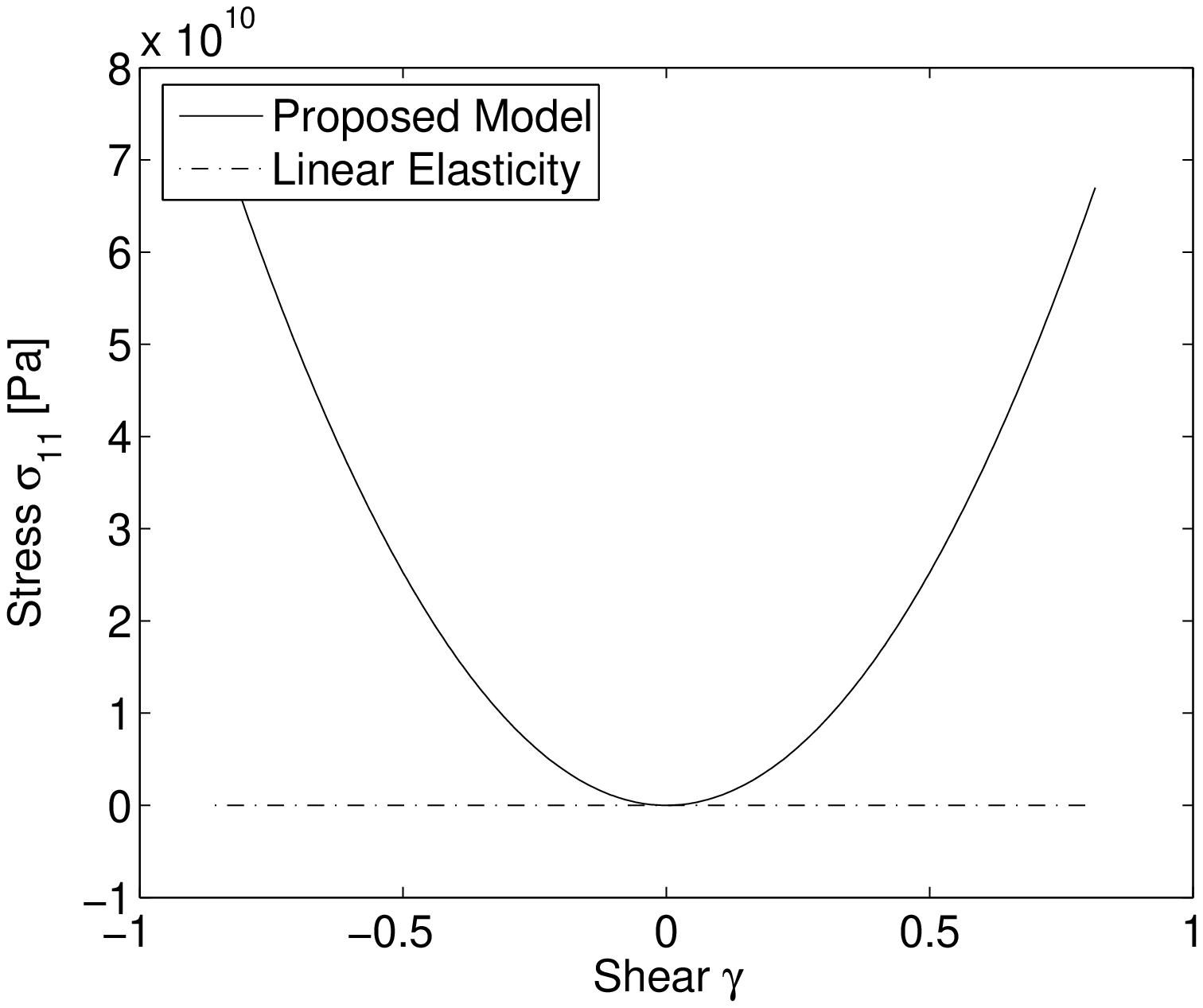}
\includegraphics[width=2.75in]{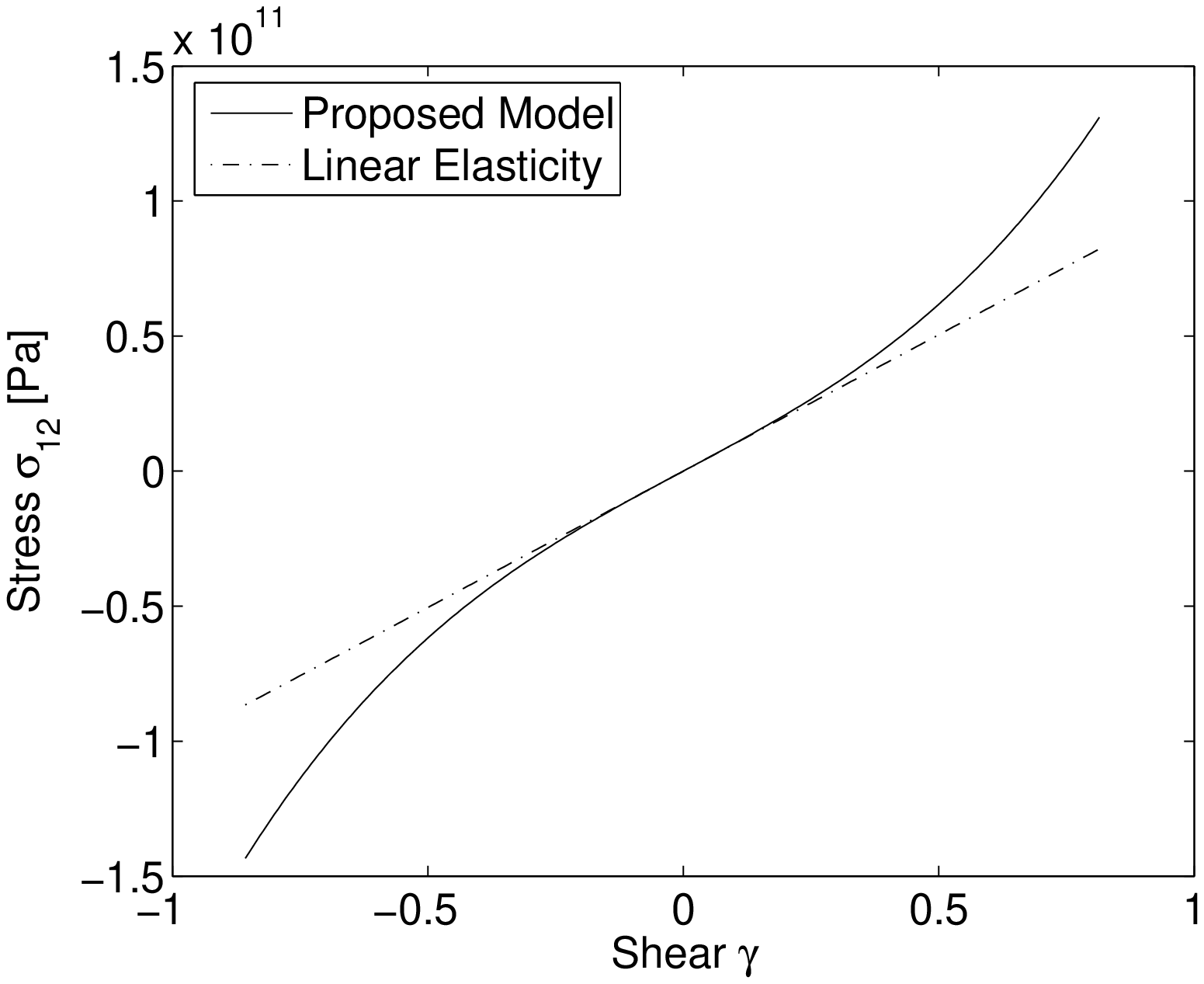}
\caption{Model response to simple shear}
\label{fig:modelcmp2}
\end{figure}

As an initial test for the model, two simple deformation gradient
tensors have been applied as inputs:
\begin{displaymath}
\mathbf{F}=\left[\begin{array}{ccc} \lambda & 0 & 0 \\ 0 & 1 & 0 \\ 0
    & 0 & 1 \end{array} \right]\;\;\;\mathrm{and}\;\;\; 
\mathbf{F}=\left[\begin{array}{ccc} 1 & \gamma & 0 \\ 0 & 1 & 0 \\ 0
    & 0 & 1 \end{array} \right] .
\end{displaymath}
Figure \ref{fig:modelcmp} displays the Cauchy stresses $\sigma$ as
functions of the stretch $\lambda$.  Similarly to linear elasticity,
the model predicts that the stresses $\sigma_{11}$ and
$\sigma_{22}=\sigma_{33}$ are increasing functions of $\lambda$.  The
apparent agreement between the proposed model and linear elasticity in
the small strains regime is not surprising given that the model
parameters are fitted to the linear elasticity coefficients at zero
strain.  The physically desirable behavior of the stresses going to
$+\infty$ ($-\infty$) as the stretch $\lambda$ goes to $+\infty$ ($0$)
is also present.  The behavior of the model in simple shear is shown
in Figure \ref{fig:modelcmp2}.  As in the previous figure, the stress
curves predicted by the model are tangent to the stress curves for
linear elasticity.  The most notable difference in this case is that
the model predicts nonzero stress $\sigma_{11}$ while according to
linear elasticity $\sigma_{11}=0$.

\section{Variational formulation and finite element discretization}
\label{sec:variational}

Consider a body $\mathcal{B}$ with boundary $\partial
\mathcal{B}=\mathcal{A}_1 \cup \mathcal{A}_2$.  The boundary
$\mathcal{A}_1$ consists of all surface points where displacement is
applied and the boundary $\mathcal{A}_2$ of all surface points where
tractions are applied ($\mathcal{A}_1 \cap \mathcal{A}_2 = \phi$). The
boundary value problem can be formulated as (following
\cite{Ball:77a}):
\begin{eqnarray}
\mathrm{Div}(\mathbf{P}) & = & 0\;\;\;\;\;\;\;\;\;\;\;\;\;\;\;\;\mathrm{in}\; \mathcal{B}, \label{eq:varequilibrium}\\
\mathbf{P} & = & \mathbf{P}(\nabla \mathbf{u},
\mathbf{x})\;\;\;\mathrm{in}\; \mathcal{B}, \label{eq:varconstitutive}\\
\mathbf{u} & = & \mathbf{\bar{u}}\;\;\;\;\;\;\;\;\;\;\;\;\;\;\;\mathrm{on}\; \mathcal{A}_1, \label{eq:varimposeddisp}\\
\mathbf{P} \mathbf{n} & = & \bar{\mathbf{t}}\;\;\;\;\;\;\;\;\;\;\;\;\;\;\;\;\mathrm{on}\; \mathcal{A}_2,\label{eq:varimposedtrac}
\end{eqnarray}
where (\ref{eq:varequilibrium}) expresses the equilibrium condition of
the solid in the absence of body forces, (\ref{eq:varconstitutive}) is
the constitutive law for the solid, (\ref{eq:varimposeddisp})
expresses the boundary condition for the section of the boundary
$\mathcal{A}_1$ on which displacement is imposed and
(\ref{eq:varimposedtrac}) refers to the section of the boundary
$\mathcal{A}_2$ on which the tractions are imposed.  For hyperelastic
materials possessing strain energy function $\psi$ the first
Piola-Kirchhoff stress tensor is given by $\mathbf{P}=\frac{\partial
\psi}{\partial \mathbf{F}}$ where $\mathbf{F}=\nabla \mathbf{u}$.

It can be shown (\cite{Ball:77a}) that the solution $\mathbf{u}$ of
the problem posed by (\ref{eq:varequilibrium}-\ref{eq:varimposedtrac})
in the case of polyconvex strain energy function $\psi$ is the
minimizer of the functional
\begin{equation}
J(\mathbf{u})=\int_{\mathcal{B}} \psi(\mathbf{x}, \mathbf{u}, \nabla
\mathbf{u}) dV - \int_{\partial \mathcal{A}_2} \mathbf{u} \cdot
\bar{\mathbf{t}} dS, \label{eq:functional}
\end{equation}
which can be used to formulate the finite element discretization.  The
spatial discretization is accomplished by representing the body
$\mathcal{B}$ as a union of disjoint elements, $\mathcal{B}=
\bigcup_{e=1}^{N_{elem}} \Omega_e$.  Even though many different
elements can be used for the discretization, for simplicity, we will
assume that the discretization has been achieved by the use of second
order tetrahedral elements.  The unknowns to be solved for are the
nodal displacements $\mathbf{u}_n$.  For each element, the displacement is
expressed as a sum of the nodal functions,
$\mathbf{u}^{(e)}=\sum_{a=1}^{10} N_a^{(e)} \mathbf{u}_a^{(e)}$.  The
global displacement approximation becomes
$\mathbf{u}_h=\sum_{e=1}^{N_{elem}} \sum_{a=1}^{10} N_a^{(e)}
\mathbf{u}_a^{(e)}$. Substitution of $\mathbf{u}_h$ into
(\ref{eq:functional}) leads to a discrete functional
$J_h(\mathbf{u}_h)$.  The minimization procedure for the discrete
functional gives rise to a system of equations which can be solved for
the nodal displacement. Further details about the finite element
procedure can be found in \cite{Radovitzky:99}.

\section{Numerical examples}
\label{sec:numerical}

In this section we will consider two basic examples which illustrate
the agreement of the model with linear theory at small strains and the
departure from it at large deformations.

\subsection{2D Example -- Plate with Hole}

\begin{figure}
\includegraphics[width=2.75in]{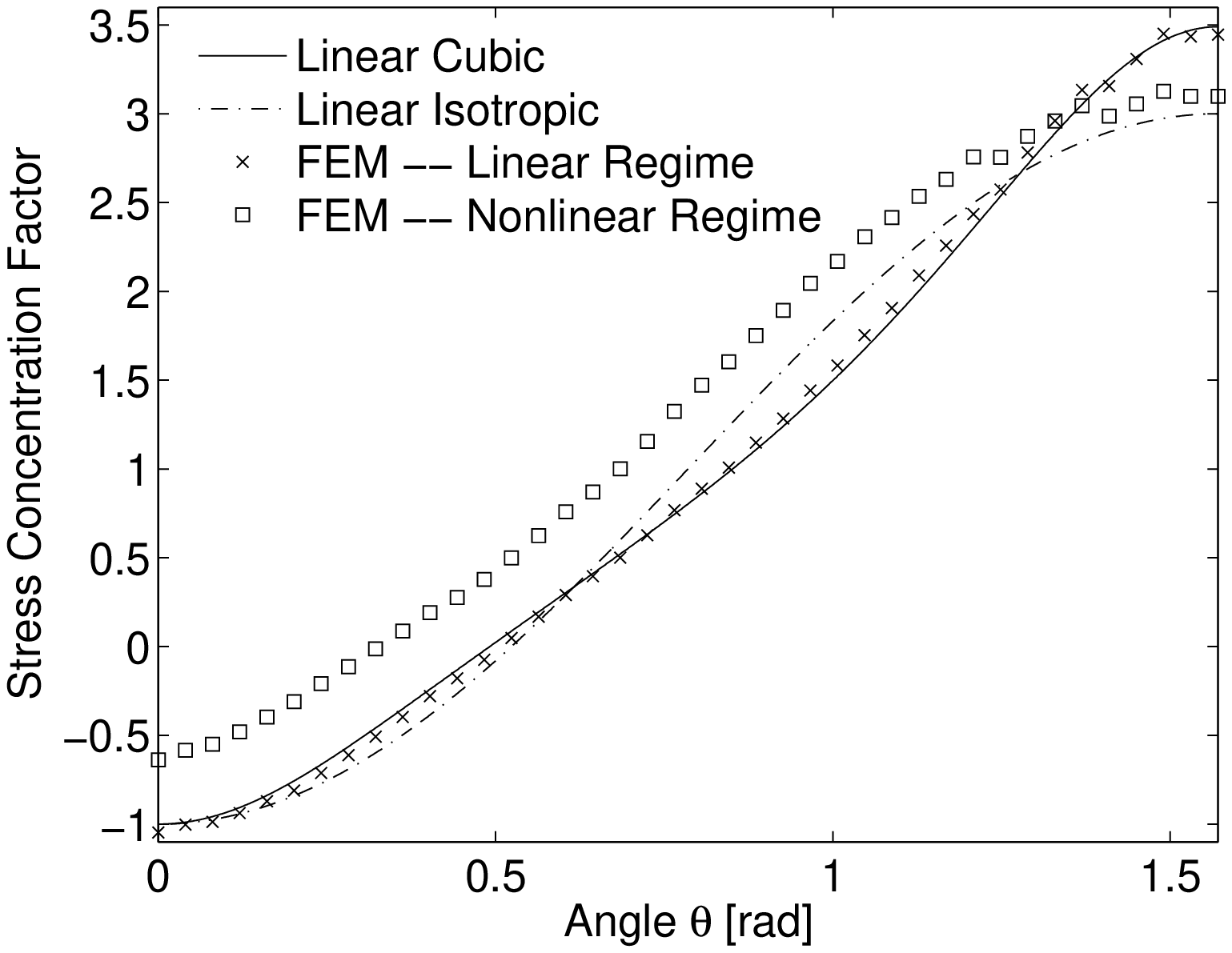}
\includegraphics[width=2.75in]{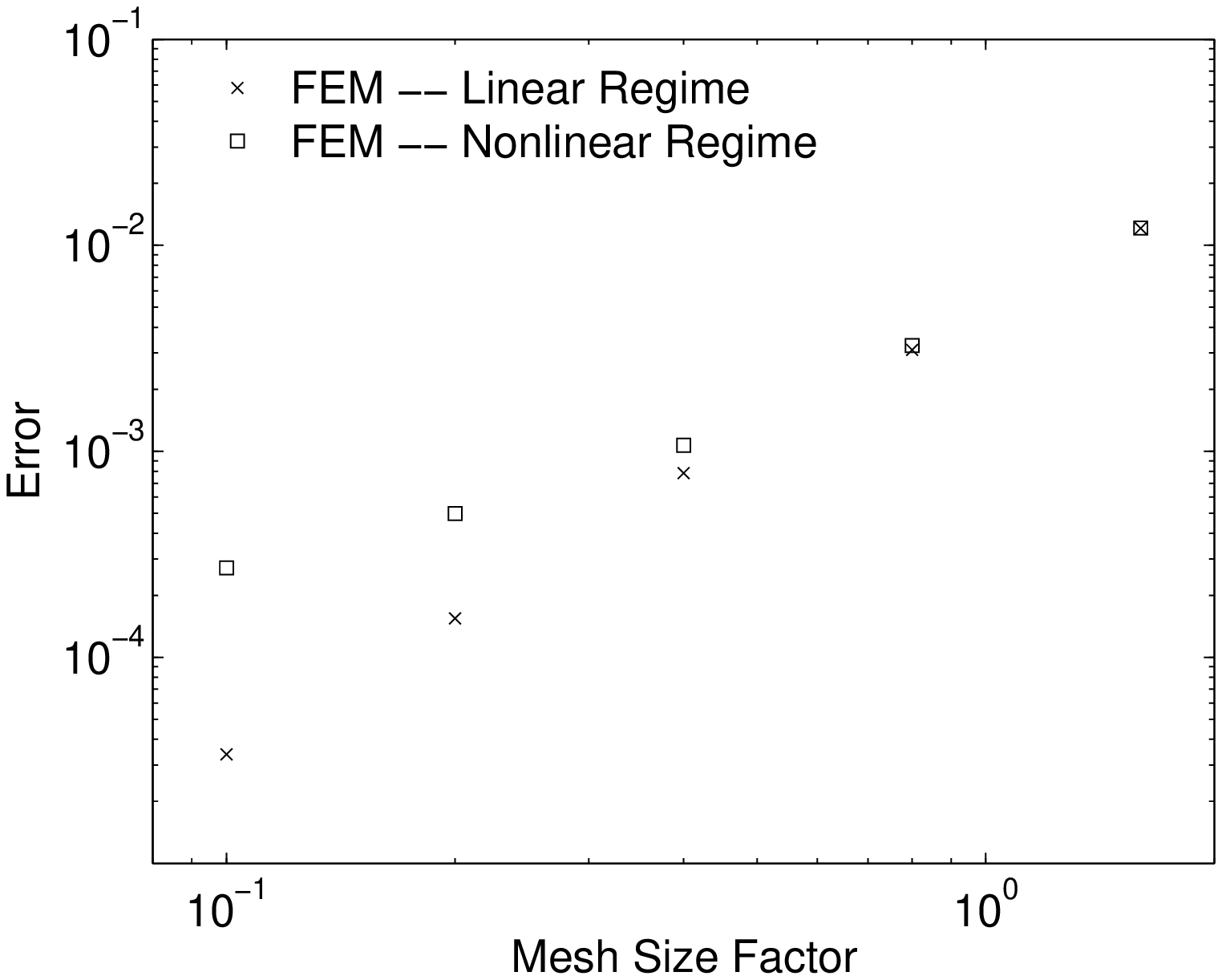}
\caption{Stress concentration and convergence plots for plate with an
initially circular hole}
\label{fig:plate-hole}
\end{figure}

To verify the consistency of our model with linear theory and to check
its convergence we considered the problem of uniaxial stress applied
to a plate with initially circular hole.  In addition to the well
known analytical solution for isotropic linear elastic material, this
problem has analytic solution for orthotropic linear elastic materials
\cite{Green:68}.  The stress concentration factor, calculated from
this solution specialized to cubic anisotropy with one symmetry axis
perpendicular to the plane of the plate and another symmetry axis
aligned with the loading direction, is shown on the left side of
Figure \ref{fig:plate-hole}.  The numerical solution obtained from our
model is in good agreement with this analytical solution when the
applied stress is small compared to the elastic constants of the
material.  As the stress is increased to become comparable to the
elastic constants the nonlinearities become important and
approximately the same reduction in the minimum and the maximum stress
concentration is observed.

The plot on the right of Figure \ref{fig:plate-hole} shows the
dependence of the error on the mesh size.  Within the linear regime
the convergence is quadratic as expected for Newton-Raphson solvers.
The convergence is somewhat reduced for the nonlinear regime, but it
is still within acceptable levels.

\subsection{3D Example -- Circular Bar}

\begin{figure}
\includegraphics[width=1.75in]{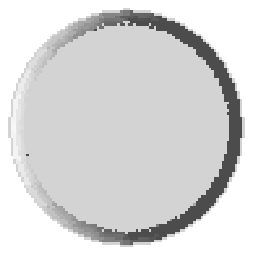}
\includegraphics[width=1.75in]{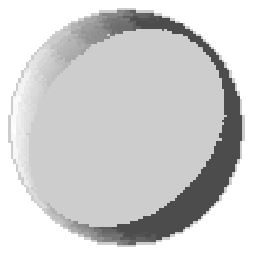}
\includegraphics[width=1.75in]{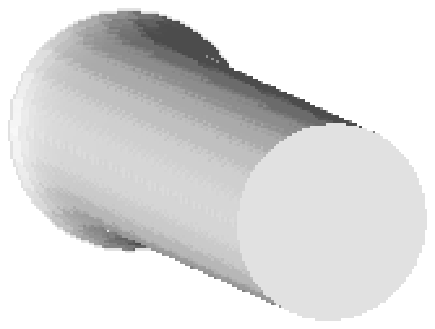}\\
\includegraphics[width=1.75in]{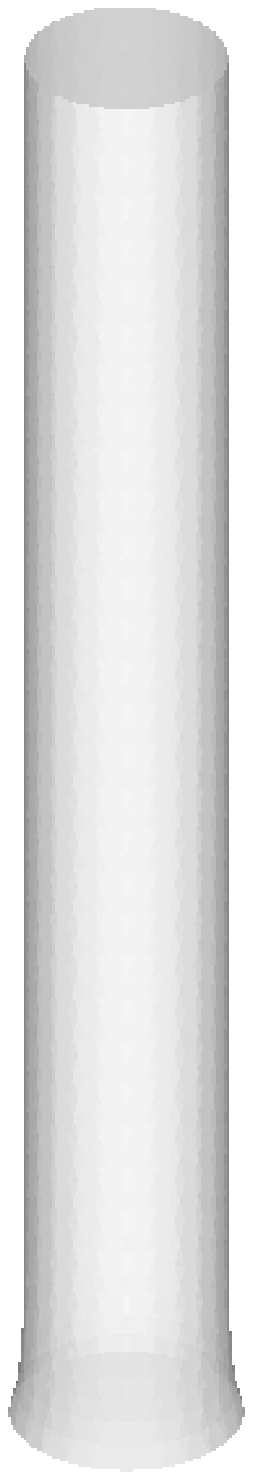}
\includegraphics[width=1.75in]{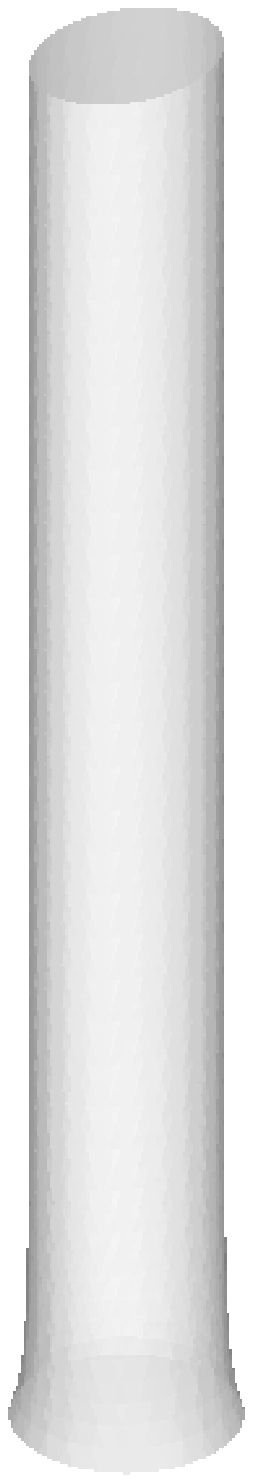}
\includegraphics[width=1.75in]{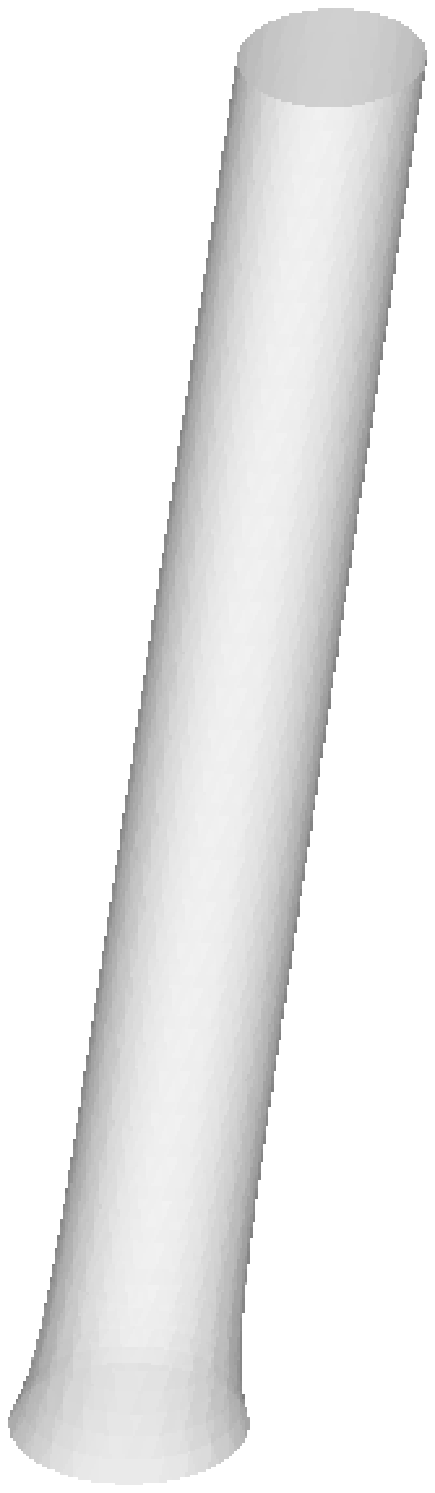}\\
\begin{verbatim}       [100]                [110]                 [122]\end{verbatim}
\caption{Response of single crystal bar to imposed displacement at
different orientations of the crystal relative to the bar axis}
\label{fig:bar-response}
\end{figure}

The problem which we will consider here is the extension of a single
crystal cylindrical bar.  The bottom end of the bar is clamped and the
side surface of the bar is traction free. The top end is displaced in
the axial direction by a specified amount, but it is left free to move
in the plane perpendicular to the original bar axis. Three different
orientations of the bar axis relative to the crystal will be
considered.  In each case the bar axis will coincide with one of the
following crystallographic directions: [100], [110] and [122].  Top
and side views of the deformed bar for the three cases are shown in
Figure \ref{fig:bar-response}. For better visualization the applied
displacement is equal to $100$\% of the bar length, but the behavior
is similar at smaller stretches.

In the first case the cross-section of the bar remains approximately
circular and the extension process is approximately symmetric about
the bar axis.  Similar behavior is observed if the bar axis is aligned
with the [111] direction.

The anisotropic response is clearly observed in the second case.  The
contraction of cross-section of the bar in the two directions is
markedly different due to the anisotropy brought by the term
containing $I_4$.  As it could be expected the contraction in the
[001] direction is much less than in the [$1\bar{1}0$] direction. A
possible interpretation of this effect for metallic lattices is that
part of the contraction in the latter direction is accomplished by
atomic bond rotation rather than pure extension/contraction of the
bonds.

The third case illustrates the development of transverse displacements
when the crystal lattice lacks enough symmetries relative to the
loading axis.  The tilting effect is completely due to the anisotropy.
If the movement of the top end is restricted significant transverse
stresses will develop.

\section{Conclusions}
\label{sec:conclusion}

A new model for materials with cubic anisotropy has been proposed in
this paper.  The model is based on additively decoupled strain energy
function which satisfy the polyconvexity condition and therefore
guarantees existence of minimizing sequences for the appropriate
variational functionals.  The polyconvexity of new strain energy terms
capturing the anisotropy of cubically symmetric systems has been
shown.  A simple strain energy function capable of capturing the
fundamental effect of the anisotropy ratio has been suggested and
tested in numerical simulations which reveal that the model possesses
many of the relevant physically desirable properties.

The main difference between this model and the orthotropic models in
the literature (for example, \cite{Scho-Neff:03}) is the use of a
single fourth order structural tensor.  In spite of the complications
coming from the higher order of the tensor, the model avoids a major
complication which the orthotropic models face: enforcing the equality
of the properties in the three mutually perpendicular symmetry
directions.


\section*{Acknowledgment}

The support of ASC through grant $ASC????????$ is gratefully
acknowledged.

\appendix

\section{Invariants derivatives}
\label{sub:Invariants-derivatives}

The computation of the stress tensor $\mathbf{S}$ and the tangent
tensor $\mathbb{C}$ requires the first and second derivatives of the
invariants.  Expressions for these derivatives deduced in the
reference configuration are given below.


\begin{itemize}
\item First derivatives
\begin{eqnarray*}
\frac{\partial I_{1}}{\partial C_{IJ}} & = & \delta_{IJ} \\
\frac{\partial I_{2}}{\partial C_{IJ}} & = & I_{1}\delta_{IJ}-C_{IJ}\\
\frac{\partial I_{3}}{\partial C_{IJ}} & = & I_{3}C_{IJ}^{-1}\\
\frac{\partial I_{4}}{\partial C_{IJ}} & = & {\displaystyle
  \frac{1}{2}} \left(M_{IJPQ}+ M_{JIPQ} + M_{PQIJ} +M_{PQJI}
  \right) C_{PQ}\\
\frac{\partial I_{5}}{\partial C_{IJ}} & = & \frac{1}{2} \left[\left
  (\delta_{IM} \delta_{JN} +\delta_{IN} \delta_{JM} \right) M_{MNPQ}
  C_{PQ}^{2} +\right.\\
 & & \left.C_{MN}M_{MNPQ} \left( \delta_{IP}C_{JQ}+C_{IQ} \delta_{JP}
  +C_{PI} \delta_{QJ}+ \delta_{IQ}C_{JP}\right)\right] \\
\frac{\partial I_{6}}{\partial C_{IJ}} & = & \frac{1}{2} \left[\left(
  \delta_{IM} C_{JN}+ \delta_{JM}C_{IN} +C_{IM}\delta_{JN} +C_{JM}
  \delta_{IN} \right) M_{MNPQ}C_{PQ}^{2}+\right. \\
& & \left.C_{MN}^{2} M_{MNPQ}\left( \delta_{IP} C_{JQ}+C_{IQ} \delta_{JP}+
C_{PI} \delta_{QJ} +\delta_{IQ} C_{JP}\right) \right] \\
\frac{\partial\bar{I}_{M}}{\partial C_{IJ}} & = & \frac{1}{2}
  \delta_{MN} M_{MNPQ}\left( \delta_{PI}\delta_{QJ} +\delta_{PJ}
  \delta_{QI} \right) \\
\frac{\partial\bar{\bar{I}}_{M}}{\partial C_{IJ}} & = & \frac{1}{2}
  \delta_{MN}M_{MNPQ} \left(\delta_{PI} C_{QJ}+\delta_{PJ} C_{QI}+
  C_{PI} \delta_{QJ}+C_{PJ} \delta_{QI}\right) \\
\frac{\partial\hat{I}_{M}}{\partial C_{IJ}} & = & \frac{1}{2} I_{3}
  \delta_{MN} M_{MNPQ}\left(2C_{PQ}^{-1}C_{IJ}^{-1} -C_{PI}^{-1}
  C_{QJ}^{-1} -C_{PJ}^{-1}C_{QI}^{-1}\right) \\
& & \left(M_{KLJP}+M_{LKJP}\right) C_{PI}+ \left(M_{KLPJ} +M_{LKPJ}
  \right) C_{PI}+ \\
& & C_{PQ}\left(M_{PQIL}\delta_{JK}+M_{PQIK}\delta_{JL}\right)+
\end{eqnarray*}


\item Second derivatives

\begin{eqnarray*}
\frac{\partial^{2}I_{1}}{\partial C_{IJ}\partial C_{KL}} & = & 0 \\
\frac{\partial^{2}I_{2}}{\partial C_{IJ}\partial C_{KL}} & = &
  \delta_{IJ} \delta_{KL}-\frac{1}{2}\left(\delta_{IK} \delta_{JL}
  +\delta_{IL} \delta_{JK}\right) \\
\frac{\partial^{2}I_{3}}{\partial C_{IJ}\partial C_{KL}} & = & I_{3}
  \left[C_{IJ}^{-1} C_{KL}^{-1} -\frac{1}{2} \left(C_{IK}^{-1}
    C_{JL}^{-1} +C_{IL}^{-1}C_{JK}^{-1}\right)\right] \\
\frac{\partial^{2}I_{4}}{\partial C_{IJ}\partial C_{KL}} & = &
  \frac{1}{4} \left[M_{IJKL}+M_{JIKL}+M_{IJLK}+M_{JILK}+\right.\\
 & &\left.M_{KLIJ}+M_{KLJI}+M_{LKIJ}+M_{LKJI}\right] \\
\frac{\partial^{2}I_{5}}{\partial C_{IJ}\partial C_{KL}} & = &
  \frac{1}{4} \left[\left(M_{IJKP}+ M_{JIKP}\right) C_{PL}+ \left(
    M_{IJPK} +M_{JIPK}\right) C_{PL}+\right. \\
& & \left(M_{IJPL} +M_{JIPL}\right) C_{PK} + \left(M_{IJLP}+M_{JILP}
  \right)  C_{PK} + \\
& & \left(M_{KLIP}+M_{LKIP} \right) C_{PJ}+\left( M_{KLPI}+ M_{LKPI}
  \right)  C_{PJ}+ \\
& & \left(M_{KLJP}+M_{LKJP}\right)
  C_{PI}+\left(M_{KLPJ}+ M_{LKPJ}\right) C_{PI}+ \\
& & C_{PQ}\left(M_{PQIL}\delta_{JK}+M_{PQIK}\delta_{JL}\right)+
  C_{PQ}\left(M_{PQIL}\delta_{JK}+M_{PQIK}\delta_{JL}\right)+ \\
& & C_{PQ} \left( M_{PQIL}\delta_{JK}+M_{PQIK}\delta_{JL}\right)+
  \left. C_{PQ}\left( M_{PQKI}\delta_{JL} +M_{PQLI}\delta_{JK} \right)
  \right] \\
\frac{\partial^{2}I_{6}}{\partial C_{IJ}\partial C_{KL}} & = &
  \frac{1}{4} \left[\left(M_{ILPQ} \delta_{JK}+M_{IKPQ} \delta_{JL}
    +M_{JLPQ} \delta_{IK}+M_{JKPQ} \delta_{IL}
    +\right. \right. \\
& & \left. +M_{KJPQ} \delta_{IL} M_{LJPQ}\delta_{IK}+M_{KIPQ}
    \delta_{JL} +M_{LIPQ} \delta_{JK}
    \right)C_{PQ}^{2}+ \\
& & C_{LP}\left(M_{KPIQ}C_{JQ}+M_{KPJQ}C_{QI}+ M_{KPQJ}C_{QI}+
    M_{KPQI} C_{QJ}\right)+\\
& & C_{LP}\left(M_{PKIQ} C_{JQ}+M_{PKJQ} C_{QI}+M_{PKQJ}
    C_{QI}+M_{PKQI} C_{QJ}\right)+\\
& & C_{KP}\left(M_{LPIQ}C_{JQ} +M_{LPJQ}C_{QI} +M_{LPQJ}C_{QI}
    +M_{LPQI} C_{QJ}\right)+\\
& & C_{KP}\left(M_{PLIQ}C_{JQ} +M_{PLJQ}C_{QI} +M_{PLQJ}C_{QI}
    +M_{PLQI} C_{QJ}\right)+\\
& & C_{PQ}^{2} \left(M_{PQIL} \delta_{JK}+M_{PQIK} \delta_{JL}
    +M_{PQJL} \delta_{IK}+ M_{PQKJ} \delta_{IL}+\right. \\
& & M_{PQLJ} \delta_{IK}+ \left.\left.M_{PQKI} \delta_{JL}+ M_{PQLI}
    \delta_{JK} +M_{PQJK} \delta_{IL} \right)\right]\\
\frac{\partial^{2}\bar{I}_{M}}{\partial C_{IJ}\partial C_{KL}} & = &
  0 \\
\frac{\partial^{2}\bar{\bar{I}}_{M}}{\partial C_{IJ}\partial C_{KL}} &
  = & \frac{1}{4} \delta_{MN}M_{MNPQ} \left[\delta_{PI}
    \left(\delta_{QK} \delta_{JL}+ \delta_{QL}\delta_{JK} \right)
    +\delta_{QI} \left(\delta_{PK}\delta_{JL}+ \delta_{PL}\delta_{JK}
    \right) \right.\\
& & \left.\delta_{PJ}\left(\delta_{IK} \delta_{QL}+ \delta_{IL}
    \delta_{QK} \right)+\delta_{QJ} \left(\delta_{IK} \delta_{PL}+
    \delta_{IL} \delta_{PK}\right)\right] \\
\frac{\partial^{2}\hat{I}_{M}}{\partial C_{IJ}\partial C_{KL}} & = &
  \frac{1}{2} I_{3} \delta_{MN}M_{MNPQ} \left[\left(2C_{IJ}^{-1}
    C_{PQ}^{-1} -C_{PI}^{-1}C_{QJ}^{-1} -C_{PJ}^{-1} C_{QI}^{-1}
    \right) C_{KL}^{-1}-\right.\\
& & C_{PQ}^{-1}\left(C_{IK}^{-1}C_{JL}^{-1} +C_{IL}^{-1}C_{JK}^{-1}
    \right) -C_{IJ}^{-1} \left(C_{PK}^{-1} C_{QL}^{-1}+ C_{PL}^{-1}
    C_{QK}^{-1} \right)\\
& & \frac{1}{2} C_{PI}^{-1} \left(C_{QK}^{-1} C_{JL}^{-1} +C_{QL}^{-1}
    C_{JK}^{-1}\right)+\frac{1}{2} C_{QJ}^{-1} \left(C_{PK}^{-1}
    C_{IL}^{-1} +C_{PL}^{-1}C_{IK}^{-1}\right)\\
& & \left.\frac{1}{2} C_{PJ}^{-1} \left(C_{QK}^{-1} C_{IL}^{-1}
    +C_{QL}^{-1} C_{IK}^{-1}\right)+ \frac{1}{2} C_{QI}^{-1}
    \left(C_{PK}^{-1} C_{JL}^{-1}+C_{PL}^{-1} C_{JK}^{-1} \right) \right]
\end{eqnarray*}

\end{itemize}

\section{Proofs of basic properties}
\label{sec:proofs}

\begin{description}
\item [Statement.]\begin{equation}
\begin{array}{l}
\begin{array}{l}
\mathbf{H}^T \mathbf{F}:{\bf
  {M}}:\mathbf{F}^{T}\mathbf{H}=\mathbf{H}^T \mathbf{F} :{\bf {M}}:\mathbf{H}^{T}\mathbf{F}\end{array}\end{array}\label{eq:A1}\end{equation}

\item [Proof.]Making use of the indicial notation we have\[
\mathbf{H}^T\mathbf{F}:{\bf
{M}}:\mathbf{F}^{T}\mathbf{H}=H_{MI}F_{MJ}M_{IJKL}F_{NK}H_{NL}.
\]
Considering the symmetries of ${\bf {M}}$, expression
(\ref{eq:M-symmetries}), and redefining properly dummy indices, the
above expression can be rewritten as
\[
\begin{array}{l}
\mathbf{H}^T\mathbf{F}:{\bf {M}}:\mathbf{F}^{T}\mathbf{H}=H_{MI}F_{MJ}M_{IJLK}F_{NK}H_{NL}=H_{MI}F_{MJ}M_{IJLK}H_{NL}F_{NK}\\
\qquad\qquad\qquad\quad\,\,\,=\mathbf{H}^T\mathbf{F}:{\bf
  {M}}:\mathbf{H}^{T}\mathbf{F}.\Box \end{array}\]

\end{description}
\begin{description}
\item [Statement.]\begin{equation}
\begin{array}{l}
\begin{array}{l}
\mathbf{F}^T\mathbf{H}:{\bf {M}}:\mathbf{H}^{T}\mathbf{F}=
\mathbf{H}^T \mathbf{F} :{\bf {M}}:\mathbf{H}^{T}\mathbf{F}\end{array}\end{array}\label{A2}\end{equation}

\item [Proof.]Analogously to the previous statement, we have\[
\begin{array}{l}
\mathbf{F}^T \mathbf{H}:{\bf {M}}:\mathbf{H}^{T}\mathbf{F}=F_{MI}H_{MJ}M_{IJKL}H_{NK}F_{NL}=F_{MI}H_{MJ}M_{JIKL}F_{NK}H_{NL}\\
\qquad\qquad\qquad\quad\,\,\,=H_{MJ}F_{MI}M_{JIKL}H_{NK}F_{NL}=\mathbf{H}^T
\mathbf{F} :{\bf {M}}:\mathbf{H}^{T}\mathbf{F}.\Box\end{array}\]

\item [Statement.]\begin{equation}
\begin{array}{l}
\begin{array}{l}
\mathbf{H}^T \mathbf{H}:{\bf
  {M}}:\mathbf{F}^{T}\mathbf{F}=\mathbf{F}^T \mathbf{F} :{\bf {M}}:\mathbf{H}^{T}\mathbf{H}\end{array}\end{array}\label{eq:A3}\end{equation}

\item [Proof.] Similarly to the two statements above, we have\[
\begin{array}{l}
\mathbf{H}^T \mathbf{H} :{\bf {M}}:\mathbf{F}^{T}\mathbf{F}=H_{MI}H_{MJ}M_{IJKL}F_{NK}F_{NL}=F_{NK}F_{NL}M_{IJKL}H_{MI}H_{MJ}\\
\qquad\qquad\qquad\quad\,\,\,=F_{NK}F_{NL}M_{KLIJ}H_{MK}H_{ML}=\mathbf{F}^T
\mathbf{F}:{\bf {M}}:\mathbf{H}^{T}\mathbf{H}.\Box\end{array}\]

\end{description}

\begin{description}
\item[Statement.] If $\mathbf{e}$ is a vector and $\mathbf{A}$,
  $\mathbf{B}$ and $\mathbf{C}$ are second order tensors then
\begin{eqnarray}
\mathbf{A}:\mathbf{B} \otimes \mathbf{B}: 
\mathbf{C} & = & (\mathbf{A}:\mathbf{B}) (\mathbf{B}:\mathbf{C})
\label{eq:ABBC} \;\;\;\mathrm{ and} \\
\mathbf{e} \otimes \mathbf{e}:\mathbf{A}^T\mathbf{A} & = &
(\mathbf{A}
\mathbf{e})^2 \label{eq:eeATA}.
\end{eqnarray}
\item[Proof.] Using index notation, we have the following equalities:
\begin{displaymath}
\mathbf{A}:\mathbf{B} \otimes \mathbf{B}: 
\mathbf{C} = A_{IJ} B_{IJ} B_{KL} C_{KL} 
= (\mathbf{A}:\mathbf{B}) (\mathbf{B}:\mathbf{C})
\end{displaymath}
and
\[
\mathbf{e} \otimes \mathbf{e}:\mathbf{A}^T\mathbf{A} = e_I e_J
A_{KI} A_{KJ} = A_{KI} e_I A_{KJ} e_J =  (\mathbf{A} \mathbf{e})^2. \Box
\]
\end{description}

\begin{description}
\item[Statement.] If $\alpha > 0$ and $p\ge2$ satisfy the condition
\begin{equation}
\frac{\alpha+1}{\alpha}\ge\frac{p}{p-1}, \label{eq:neff-lemma1}
\end{equation}
then the function $\phi(x,y)=\frac{x^p}{y^\alpha}$ is convex when
considered as a function of both arguments simultaneously.
\item[Proof.] The complete proof is given in \cite{Scho-Neff:03}.  We
will only note here that the condition (\ref{eq:neff-lemma1}) is
equivalent to the positive definiteness of the second derivative (the
Hessian) of $\phi$. $\Box$
\end{description}

\section{Characteristic polynomial}
\label{sec:char_polynomial}

The characteristic polynomial, also known as Cayley-Hamilton polynomial,
for a matrix ${\bf {C}}\in M^{3\mathrm{x}3}$, is the equation resulting from
the eigenvalue problem corresponding to that matrix ${\bf {C}}$,
i.e., $\mathrm{det}(\lambda{\bf {1}-{\bf {C}}})=0$, giving
\[
0=\lambda^{3}-\mathrm{tr}({\bf {C}})\lambda^{2}+\mathrm{tr}
(\mathrm{adj} ({\bf {C}}))\lambda- \mathrm{det}({\bf {C}}),\]
or equivalently
\begin{equation}
0={\bf {C}}^{3}-\mathrm{tr} ({\bf {C}}){\bf {C}}^{2}+\mathrm{tr}
(\mathrm{adj} ({\bf {C}})){\bf {C}}-\mathrm{det} ({\bf {C}}){\bf
  {1}}. \label{eq:cayley-hamilton}
\end{equation}

\bibliographystyle{elsart-num}
\bibliography{manuscript}
\end{document}